\documentclass{aa}  

\usepackage{graphicx}
\usepackage{txfonts}
\usepackage{newtxtext,newtxmath}
\usepackage[T1]{fontenc}
\usepackage{mathtools}
\usepackage{amsmath}
\usepackage{amssymb}	
\usepackage{makecell, multirow}
\usepackage{lscape}
\usepackage{subcaption}
\usepackage{float}
\usepackage[section]{placeins} 
\usepackage{threeparttable}
\usepackage{placeins}
\usepackage[usenames, dvipsnames]{color}

\begin{document}

   \title{Water content of rocky exoplanets in the habitable zone}

      \author{Ádám Boldog\inst{1,2,3}\fnmsep\thanks{E-mail: boldog.adam@csfk.org}, 
      Vera Dobos\inst{4,3},
      László L. Kiss\inst{1,2,5}, 
      Marijn van der Perk\inst{4}
      \and Amy C. Barr\inst{6}}

   \institute{Konkoly Observatory, HUN-REN Research Centre for Astronomy and Earth Sciences, \\ Konkoly Thege Miklós út 15-17, H-1121 Budapest, Hungary
         \and
             CSFK, MTA Centre of Excellence, Budapest, Konkoly Thege Miklós út 15-17, H-1121, Hungary
        \and
        MTA-ELTE Exoplanet Research Group, 9700 Szombathely, Szent Imre h. u. 112, Hungary
        \and
        Kapteyn Astronomical Institute, University of Groningen, 9747 AD, Landleven 12, Groningen, The Netherlands
        \and
        ELTE Eötvös Loránd University, Institute of Physics, Pázmány Péter sétány 1/A, 1117
        \and
        Planetary Science Institute, 1700 E. Ft. Lowell, Suite 106, Tucson, AZ 85719, USA
             }

   \date{Received ; accepted }

 
  \abstract
   {In this study we investigated the interiors of rocky exoplanets in order to identify those that may have large quantities of water. 
We modelled the interiors of 28 rocky exoplanets, assuming four different layers - an iron core, a rock mantle, a high-pressure ice layer, and a surface ice/water layer. Due to observational bias, our study is limited to habitable zone exoplanets. We determined a range of possible water mass fractions for each planet consistent with the modelled planetary structures. We calculated the tidal heating experienced by these exoplanets through gravitational interactions with their host stars, assuming a temperature- and composition-dependent Maxwell viscoelastic rheology. Assuming radioactive elemental abundances observed in Solar System meteorites, we also calculated the radiogenic heat flux inside the planets. We estimated the probability of the presence of a thick ocean layer in these planets, taking into account the effect of both tidal and radiogenic heating. Our results showed that Proxima~Centauri~b, Ross~128~b, Teegarden's~b and c, GJ~1061~c and d, and TRAPPIST-1~e may have an extended liquid water reservoir. Furthermore, extremely high H$_2$O-content of the exoplanets Kepler-62~f, Kepler-1652~b, Kepler-452~b, and Kepler-442~b suggests that these planets may maintain a water vapour atmosphere and may in fact be examples of larger ocean worlds. Upon the discovery of new rocky exoplanets beyond the habitable zone, our study can be extended to icy worlds.}
   
   
   

    \titlerunning{Water content of rocky exoplanets}
    \authorrunning{Boldog et al.}
   \maketitle
%

\section{Introduction}

According to the Planetary Habitability Laboratory\footnote{ http://phl.upr.edu/projects/habitable-exoplanets-catalog} (PHL), to date there are 59 known rocky exoplanets that are orbiting in the habitable zones (HZs) of their host stars. 
There are different approaches to investigating the habitability of exoplanets. Since habitability on the surface is strongly dependent on the presence of an atmosphere, detecting and measuring the composition of exoplanetary atmospheres - which is in the scope of the recently launched James Webb Space Telescope \citep{greene2019, birkmann2022} - is one of the most important tools for studying the habitability of exoplanets today. However, in the absence of an atmosphere, surface life is unlikely to develop \citep{Lammer2009}. Atmospheric loss processes can result in planets that are completely deprived of their atmospheres. These processes could be particularly significant for exoplanets orbiting close to M dwarf stars. Nevertheless, planets without a significant atmospheric layer can still provide the conditions necessary for the development of life. Such planets could be habitable given the presence of an ocean under the surface, as in the case of Enceladus. Therefore, it is important to explore other possible ways of studying the habitability of exoplanets that do not rely on the presence of an atmosphere.

Relatively little attention has been paid to the possibility that life might flourish in subsurface liquid water oceans on exoplanets. In the presence of water beneath a layer of surface ice, the interior of a planet might be an ideal place for life to develop, because the surface ice layer could provide protection against energetic particles \citep{Paranicas2009, Pavlov2019}. Since water can exist beneath the surface even in the absence of an atmosphere, investigating the internal structures of exoplanets may open a new window for habitability studies that does not depend on atmospheric considerations. Most of the currently known rocky exoplanets orbit close to their stars, which likely prevents them from retaining their water. Habitable zone exoplanets are ideal targets since it is possible to investigate both surface and subsurface habitability. High H$_2$O content, for example, may imply either subsurface water reservoirs or, in the presence of an Earth-like atmosphere, global surface oceans. It is important to emphasize that investigating the possibility of subsurface water reservoirs does not exclude the presence of planetary atmospheres or liquid oceans on the surface, but extends possible habitable regions to subsurface regimes. Furthermore, investigating the interior structure of exoplanets may contribute to improving our understanding of the relationship between planetary formation and composition as well as to helping constrain models of planetary evolution \citep{Valencia2006, ElkinsTanton2008, Grasset2009}.

However, determining the interior structure of an exoplanet is a significant challenge, because the only available relevant information is either the radius or the mass of the planet or in some cases both of these parameters. Uncertainties in the planetary masses and radii can result in a wide range of possible values for density. Considering that there is a wide variety of interior structures that correspond to the same density, these uncertainties in the planetary parameters result in an extended range of plausible scenarios for the internal composition (e.g., \citealt{Rogers2010, Unterborn2015, Barr2018, Dobos2019}). Precise measurements of planetary masses and radii could narrow this range and improve our understanding of planetary interiors. 

The habitability of planet interiors depends on the availability of liquid water. Icy moons in our Solar System, such as Enceladus or Europa, are examples of rocky bodies with an extended ocean underneath a thick layer of ice \citep{Khurana1998, Zimmer2000, Iess2014, vanHoolst2016}. In the case of these moons, the oceans remain liquid due to the heat flux provided by radiogenic heating and tidal dissipation \citep{Nimmo2007}. The recently launched ESA spacecraft, the Jupiter Icy Moons Explorer (Juice) is the next step towards studying the interior of icy bodies \citep{GRASSET2013a}. Juice will investigate the icy moons of Jupiter in order to gain a deeper understanding of the interior structure, formation, and potential habitability of these bodies. Analogous to these icy moons in our Solar System, H$_2$O rich exoplanets with strong internal heating may maintain a layer of liquid water under the planetary surface. It is important to note that a substantial H$_2$O-reservoir may indicate a surface ocean or ice layer, depending on the surface temperature of the planet. Even in the case of HZ exoplanets, in order to gain more insight into the habitability of these bodies, it is crucial to gain knowledge about the water content of the planets.

For close-in exoplanets with non-zero eccentricities, tidal dissipation is caused by the rising and lowering of a tidal bulge on the planet.  The rising and lowering of the bulge during the course of the planet's orbit is resisted by the planet's own internal friction.  This results in heat \citep{CRPTidal}.  The time variability of the tidal forces exerted by the host star on the planet can be a significant contributor to the total energy budget of the planet and may play an important role in habitability as well \citep{FerrazMello2008, Jackson2008a, Jackson2008c, Behounkova2011}.

Another process that may contribute to internal heating is the decay of long-lived radioactive isotopes. Radiogenic heating is a significant contributor to the internal energy budget of the Earth \citep{MantleConvection}. \citet{Kami2011} found the contribution of radiogenic heating to the heat flux of the Earth to be $20.0_{-8.6}^{+8.8\smash[t]{\mathstrut}}$~TW, which converts to a surface heat flux of around 0.04 $\mathrm{W}/\mathrm{m}^2$. They concluded that the decay of radioactive elements is responsible for approximately half of the total internal heat flux of the Earth. 

The amount of radiogenic heating is strongly dependent on the planetary composition, in terms of the mass fraction of rock and the abundances of radioactive isotopes, chiefly $^{238}$U, $^{235}$U, $^{232}$Th, and $^{40}$K.  The radiogenic abundances in exoplanets may be quite different from what we see in our own Solar System, and could depend on the metallicities of the host star. \citet{Unterborn2015} found that the Th-abundance of solar analogues are generally higher than the solar value, suggesting that exoplanets orbiting these stars can experience higher heat fluxes from the decay of radioactive elements than planets in our Solar System. Similarly, \citet{Botelho2018} found that the Th abundance in the galactic thin disk was on average super-solar even for earlier times during the evolution of our Galaxy, and that the Sun is deficient of Th when compared to solar twin analogues. 


Here, we present our studies on the internal structures, water content, and heat budgets of a selected sample of well-characterized exoplanets. In Section 2 we describe our approach for modelling the interior structures of exoplanets and our methods for computing tidal and radiogenic heating. In Section 3 we present our results and assess the possibility of the existence of liquid water reservoirs in the investigated HZ exoplanets. We summarise our main conclusion in Section 4.

\section{Methods}

\subsection{Physical and orbital parameters}

We studied HZ rocky exoplanets to assess their habitability based on their structures and internal heat production. The TRAPPIST-1 planets were ideal candidates for interior modeling because their radii and masses are known to a high precision. We also included other HZ rocky bodies in our research since they are of key importance in habitability studies (see Table \ref{tab:measurements} for the full list). Rocky exoplanets that are orbiting far from their host stars outside of the HZ could be considered for investigation from the aspect of subsurface water reservoirs. However, only a handful of such planets are known, and all of them were discovered by gravitational microlensing (e.g. \citealt{Gould2014}; \citealt{Jung2019}; \citealt{Zang2021}). Due to the small chances of further characterisation, these planets are not considered in this study. Although the number of presumed rocky HZ exoplanets is larger than the number of exoplanets investigated in this research, we only modelled those bodies that had a probability of over 5\% of having a rocky composition (see Sect. 3.4).  To determine this, we used the Forecaster model, which uses an empirical relationship between the mass and the radius that depends on the composition of the planet (i.e. rocky or Neptunian) to predict either the mass or radius of an exoplanet for which only one of those values is known  \citep{Chen_Kipping_2016}. 

The measured masses and radii of these planets with their respective errors are shown in Table~\ref{tab:measurements}. In order to account for many possible scenarios for the interior structure, 10000 mass-radius pairs were generated for each planet based on the measured parameters and their errors. For symmetric errors, that is upper and lower errors that differed by less than 20\% (i.e. $(\mid \Delta_+ - \Delta_-\mid) / (\frac{1}{2}(\Delta_+ + \Delta_-))\leq 0.2 )$, we applied the same method as \citet{Chen_Kipping_2016}, who generated random values for masses and radii assuming a Gaussian distribution, with the mean of the errors defined as the standard deviation. The reported masses or radii of all the planets investigated in this study had symmetric errors. Unfortunately, for most of the planets either the radius or the mass was unknown. For these planets, the Forecaster model was used to estimate the planet masses and radii. Since the relation used to predict the missing parameter differed for rocky and Neptunian exoplanets, only mass-radius pairs that implied a rocky composition were used as input parameters for interior modelling. In this study, only those exoplanets that had a probability of more than 5\% of having a rocky composition -- in other words, where at least 500 of the resulting mass-radius pairs suggested rocky planets -- were considered. This narrowed our sample down to a total of 28 exoplanets from the initial 59 HZ rocky planets based on the data from PHL.

{\renewcommand{\arraystretch}{1.2}
\begin{table*}
    \centering
        \caption{Physical and orbital parameters of the investigated planets. Planets with only $m\mathrm{sin}i$ measurements are marked with an asterisk.}
        \setlength\tabcolsep{1.9pt}
    \begin{tabular}{c c c c c c c c}
         \hline
         Planet's name & Mass, $\mathrm{M}_\oplus$ & Radius, $\mathrm{R}_\oplus$ & $a$, AU & $e$ & $\mathrm{T}_{\mathrm{eq}}$, K & Stellar mass, $\mathrm{M}_\odot$ & References\\
         \hline
         \noalign{\smallskip}
         GJ 1061 c* & 1.74${\pm0.1907}$ & $\mathit{1.23\pm0.28}$ & 0.035$\pm0.001$ & <0.29 & & 0.12$\pm0.01$ & 1 \\
         GJ 1061 d* & 1.57$_{-0.25}^{+0.27}$ & $\mathit{1.17\pm0.25}$ & 0.052$\pm0.001$ & <0.54 & & 0.12$\pm0.01$ & 1 \\
         GJ 273 b* & 2.193${\pm0.953}$ & $\mathit{1.38\pm0.52}$ & 0.091$\pm0.01$ & 0.1$_{-0.03}^{+0.25}$ & & 0.29 & 2 \\
         GJ 667 C e* & 2.70$_{-1.40}^{+1.59}$ & $\mathit{1.57\pm0.69}$ & 0.213$\pm0.02$ & 0.02$_{-0.02}^{+0.004}$ & & 0.33 & 3 \\
         GJ 667 C f* & 2.70$_{-1.21}^{+1.40}$ & $\mathit{1.54\pm0.66}$ & 0.156$_{-0.017}^{+0.014}$ & 0.03$_{-0.03}^{+0.16}$ & & 0.33 & 3 \\
         K2-288 B b & $\mathit{5.52\pm3.58}$ & 1.906$\pm0.30267$ & 0.164$\pm0.03$ & & 226$\pm22$ & 0.33$\pm0.02$ & 4 \\
         K2-3 d & $\mathit{3.76\pm2.54}$ & 1.513 $\pm0.235$ & 0.2086$\pm0.01$ & 0.045$\pm0.045$ & 282$\pm24$ & 0.6$\pm0.09$ & 5 \\
         K2-72 e & $\mathit{2.63\pm1.74}$ & 1.289 $\pm0.1345$ & 0.106$_{-0.13}^{+0.09}$ & 0.11$_{-0.09}^{+0.2}$ & 280  & 0.27$_{-0.09}^{+0.08}$ & 6 \\
         Kepler-1229 b & $\mathit{3.22\pm1.99}$ & 1.40125$_{-0.13452}^{+0.1121}$ & 0.3006$_{-0.0091}^{+0.0069}$ & & & 0.5593$\pm0.0207$ & 7 \\
         Kepler-1649 b & $\mathit{1.30\pm0.87}$ & 1.02$\pm0.056$ & 0.0514$\pm0.0028$ & & 307$\pm26$ & 0.2$\pm0.01$ & 8 \\
         Kepler-1652 b & $\mathit{4.20\pm2.61}$ & 1.60$\pm0.18$ & 0.1654$_{-0.0075}^{+0.0042}$ & & 268$\pm20$ & 0.4$_{-0.05}^{+0.04}$ & 9 \\
         Kepler-186 f & $\mathit{1.96\pm1.27}$ & 1.166$\pm0.078$ & 0.432$_{-0.053}^{+0.171}$ & 0.04$_{-0.04}^{+0.07}$ & & 0.54$\pm0.02$ & 10 \\
         Kepler-296 e & $\mathit{1.64\pm1.24}$ & 1.09$_{-0.15}^{+0.12}$ & 0.169$_{-0.028}^{+0.029}$ & 0.165$\pm0.165$ & 337.0$\pm17.5$ & 0.5$_{-0.09}^{+0.07}$ & 7, 19 \\
         Kepler-296 f & $\mathit{2.20\pm1.51}$ & 1.21$_{-0.15}^{+0.14}$ & 0.255$_{-0.042}^{+0.043}$ & 0.18$\pm0.15$ & 274$\pm15$ & 0.5$_{-0.09}^{+0.07}$ & 7, 19 \\
         Kepler-442 b & $\mathit{3.21\pm1.95}$ & 1.395$_{-0.096}^{+0.101}$ & 0.409$_{-0.209}^{+0.06}$ & 0.04$_{-0.08}^{+0.04}$ & & 0.61$\pm0.03$ & 11 \\
         Kepler-452 b & $\mathit{3.78\pm2.29}$ & 1.511$_{-0.131}^{+0.143}$ & 1.046$_{-0.015}^{+0.019}$ & 0.035$\pm0.75$ & 265$\pm13$ & 1.04$\pm0.05$ & 11, 20 \\
         Kepler-62 e & $\mathit{4.36\pm2.47}$ & 1.61$\pm0.05$ & 0.427$\pm0.004$ & 0.13 & 270$\pm15$ & 0.69$\pm0.02$ & 12 \\
         Kepler-62 f & $\mathit{3.29\pm1.94}$ & 1.41$\pm0.07$ & 0.718$\pm0.007$ & 0.0943 & 208$\pm11$ & 0.69$\pm0.02$ & 12 \\
         Proxima Centauri b* & 1.17$\pm0.086$ & $\mathit{1.07\pm0.15}$ & 0.049$\pm0.002$ & 0.1$_{-0.0}^{+0.35}$ & 234$_{-14}^{+6}$  & 0.12$\pm0.015$ & 13 \\
         Ross 128 b* & 1.4$\pm0.21$ & $\mathit{1.13\pm0.21}$ & 0.0496$\pm0.0017$ & 0.116$\pm0.097$ & 256$\pm45$ & 0.168$\pm0.017$ & 14 \\
         Teegarden's b* & 1.04$\pm0.13$ & $\mathit{1.03\pm0.13}$ & 0.0252$\pm0.0009$ & 0.0$\pm0.16$ & & 0.089$\pm0.009$ & 15 \\
         Teegarden's c* & 1.11$\pm0.16$ & $\mathit{1.05\pm0.14}$ & 0.0443$\pm0.0015$ & 0.0$\pm0.16$ & & 0.089$\pm0.009$ & 15 \\
         TOI-700 d & $\mathit{1.84\pm1.13}$ & 1.144$_{-0.061}^{+0.062}$ & 0.163$\pm0.015$ & 0.032$_{-0.023}^{+0.054}$ & 295$\pm55$  & 0.416$\pm0.01$ & 16 \\
         TOI-1452 b & 4.8 $\pm1.3$ & 1.67$\pm0.07$ & 0.061$\pm0.003$ & & 326$\pm7$ & 0.249$\pm0.008$ & 17\\
         TRAPPIST-1 d & 0.388$\pm0.012$ & 0.788$_{-0.01}^{+0.011}$ & 0.02227$\pm0.00019$ & 0.00563$\pm0.00172$ & 288.0$\pm5.6$ & 0.0898$\pm0.0023$ & 18, 21 \\
         TRAPPIST-1 e & 0.692$\pm0.022$ & 0.92$_{-0.012}^{+0.013}$ & 0.02925$\pm0.00025$ & 0.00632$\pm0.0012$ & 251.3$\pm4.9$ & 0.0898$\pm0.0023$ & 18, 21 \\
         TRAPPIST-1 f & 1.04$\pm0.03$ & 1.045$_{-0.012}^{+0.013}$ & 0.03849$\pm0.00033$ & 0.00842$\pm0.0013$ & 219.0$\pm4.2$ & 0.0898$\pm0.0023$ & 18, 21 \\
         TRAPPIST-1 g & 1.32$\pm0.038$ & 1.13$_{-0.013}^{+0.015}$ & 0.04683$\pm0.0004$ & 0.00401$\pm0.00109$ & 198.6$\pm3.8$ & 0.0898$\pm0.0023$ & 18, 21 \\
         \noalign{\smallskip}
         \hline
    \end{tabular}
    \begin{tablenotes}
        \item[] Values of planetary masses, radii semi-major axes, and eccentricities, along with the mass of the host star and the estimated planetary equilibrium temperature where available. Average masses and radii values with the respective standard deviations predicted by the Forecaster code are shown in italics.
        \item[*] References: 1: \citet{Dreizler2020}, 2: \citet{HARPS2017}, 3: \citet{Anglada-Escude2013}, 4: \citet{Feinstein2019}, 5: \citet{Sinukoff2016}, 6: \citet{Dressing2017}, 7: \citet{Morton2016}, 8: \citet{Vanderburg2020}, 9: \citet{Torres2017}, 10: \citet{Torres2015} 11: \citet{Berger2018}, 12: \citet{Borucki2013}, 13: \citet{Anglada-Escude2016}, 14: \citet{Bonfils2018}, 15: \citet{CARMENES2019}, 16: \citet{Rodriguez2020}, 17: \citet{Cadieux2022}, 18: \citet{Agol2021}, 19: \citet{Cartier2015}, 20: \citet{Jenkins2015}, 21: \citet{Gillon2017}
    \end{tablenotes}
    \label{tab:measurements}
\end{table*}

The values of planetary eccentricities and semi-major axes were also randomized similarly. The same approach was used in the randomization of these parameters, as described above. In some cases, where only the upper limit of the eccentricity was known, we assumed symmetric errors centred on the measured value, with the upper limit taken as the standard deviation. Only those scenarios that resulted in $e \geq0$ were considered. For each run, all of these parameters were randomly generated in a Monte Carlo approach, which resulted in 10000 distinct combinations of different masses, radii, eccentricities, and semi-major axes for each planet. 

\subsection{Interior structure modelling}

In order to estimate the interior structures of these exoplanets, we used the model described by \citet{Dobos2019}, which is a modified version of the model of \citet{Barr2018}. In this model planets are assumed to consist of four different layers: an iron core, a rocky mantle, a layer composed of high pressure ice polymorphs (HPP), and a surface ice/water layer. The iron core is assumed to have a uniform density, while the mantle is characterised by compressed Bridgmanite (MgSiO$_3$). Above pressures of 209MPa, H$_2$O is represented as an HPP layer and is assumed to be a mixture of different ice polymorphs from ice II to VII. We used the posterior distributions generated with the Forecaster model where either the planetary mass or radius was unknown. Here we provide a short description of the interior structure model.

For each of the mass-radius pairs, internal compositions that satisfy the following equations were computed: 

\begin{align}\label{eq_phi}
            \Phi_{\mathrm{I}} + \Phi_{\mathrm{HPP}} + \Phi_{\mathrm{rock}} + \Phi_{\mathrm{Fe}} = 1, \\
                \Phi_{\mathrm{I}}~\rho_{\mathrm{I}} + \Phi_{\mathrm{HPP}}~\rho_{\mathrm{HPP}} + \Phi_{\mathrm{rock}}~\rho_{\mathrm{rock}} + \Phi_{\mathrm{Fe}}~\rho_{\mathrm{Fe}} = \rho,
        \label{eq_phi_rho}
\end{align}

where $\Phi_{\mathrm{I}}$, $\Phi_{\mathrm{HPP}}$, $\Phi_{\mathrm{rock}}$, and $\Phi_{\mathrm{Fe}}$ are the volume fractions of the surface ice layer, the HPP layer, the rocky mantle, and the iron core, respectively, while $\rho_{\mathrm{I}}$, $\rho_{\mathrm{HPP}}$, $\rho_{\mathrm{rock}}$, and $\rho_{\mathrm{Fe}}$ denote the densities of the same components, respectively. First, the thickness of the ice~I layer was estimated. This was achieved by calculating the maximum depth of the surface ice layer ($z_{\mathrm{I}}$) as 
\begin{equation}
    z_{\mathrm{I}} = P_\mathrm{I}/(\rho_{\mathrm{I}}g),
    \label{eq_ice_thick}
\end{equation}
where $P_\mathrm{I} = 209 \mathrm{MPa}$ is the pressure at which the phase transition of ice~I to high-pressure polymorphs occurs, corresponding to a temperature of $\mathrm{T}=253\mathrm{K}$ at the base of the ice I layer. $\rho_{\mathrm{I}}=1000$kg$/$m$^3$ is the density of the ice layer and $g=GM_{\mathrm{pl}}/R_{\mathrm{pl}}^2$ is the surface gravity. Once the volume fraction of the ice I layer was determined, the mean density of the remaining components was calculated. Since no measurements of the sizes of the mantle, iron core, or possible ice layers are available for these planets, the volume fractions of the remaining layers were treated as free parameters.  Given the absence of a more thorough understanding of the interior of the modelled exoplanets, uniform probability distribution was assumed for all internal structures in the model. The relative sizes of the HPP layer, rocky mantle, and the iron core were estimated such that their volume fractions satisfied Eq.\ref{eq_phi} and \ref{eq_phi_rho}. Given the large error bars in the planetary masses and radii (and therefore the densities of the planets), our modelling results in large numbers of possible interior structures with equal probabilities.

\subsection{Tidal heating}

In order to examine the phase of the ice layer, in other words whether the underlying HPP layer could be in liquid form, we took into account the effect of different heat sources. We calculated the internal heat production of the planets due to heat arising from tidal effects of the host star and due to the decay of radioactive elements. To calculate the planet's response to tidal forces, a Maxwell viscoelastic rheology was assumed, characterized by a volume-averaged rigidity and viscosity, as described by \citet{Dobos2019}.  Because of the large uncertainties in the planetary parameters, this approach is more preferable in this present study, since the applied rheology has fewer parameters than more complex rheologies, which yield slightly different behaviors \citep{Henning2009, Efroimsky2012}. The tidal heat flux was calculated by following the formula of \citet{Segatz1988}:
\begin{equation}
     F_{\mathrm{tidal}} = \frac{21}{8\pi R^2}\mathrm{Im}(k_2^*)\frac{R^5\omega^5e^2}{G}.
     \label{eq_tidal}
\end{equation} 
Here $R$ is the planetary radius, $\omega = 2\pi/P$ is the orbital frequency, $e$ is the eccentricity, $G$ is the gravitational constant, and Im($k_2^*$) is the imaginary part of the Love number. This expression is only valid for synchronously rotating bodies with non-zero eccentricities. Assuming a Maxwell viscoelastic rheology, Im($k_2^*$) was calculated as
\begin{equation}
     \mathrm{Im}(k_2^*)=\frac{57\eta\omega}{4\rho gR\biggl[1+\biggl(1+\frac{19\mu}{2\rho gR}\biggr)^2\frac{\eta^2\omega^2}{\mu^2}\biggr]},
\end{equation}
where $\rho$ is the density, $\mu$ is the volume-averaged shear modulus, and $\eta$ is the volume-averaged viscosity. In each layer, the values of both $\mu$ and $\eta$ depend on the state of the given material. As the temperature rises, the fraction of melted material in the layer increases. In response, the viscosity and rigidity of the material decreases. Therefore, both the viscosity and the shear modulus are strongly dependent on temperature. Expressions for the shear moduli and viscosity values of the different layers can be found in \citet{Barr2018}. \\

For the viscoelastic properties of the rock layer, the model described in \citet{DobosTurner2015} was used, which is based on previous works of \citet{FischerSpohn1990}, \citet{Moore2003}, and \citet{Henning2009}. Here, the tidal response of rock is determined by three threshold temperatures, the solidus temperature ($T_\mathrm{s}=1600\mathrm{K}$), above which the rock becomes partially melted, the breakdown temperature ($T_\mathrm{b}=1800\mathrm{K}$), where the volume fraction of solid rock crystal and melted rock are equal, and the liquidus temperature ($T_\mathrm{l}=2000\mathrm{K}$), above which the rock is completely melted. This results in four different regimes for the rock layer (below $T_\mathrm{s}$, between $T_\mathrm{s}$ and $T_\mathrm{b}$, between $T_\mathrm{b}$ and $T_\mathrm{l}$, and above $T_\mathrm{l}$), where the shear modulus and viscosity are expressed differently. A more detailed description of the behavior of the rock layer can be found in \citet{Barr2018}. For the HPP and ice I layers, the formulas for viscosity and rigidity also vary based on the state of the layer. If the temperature of the ice layers exceeded the melting point, both $\mu$ and $\eta$ took on the properties of liquid water ($\mu = 0$ ,  $\eta=10^{-3}$~Pa~s). \\

In order to calculate the response of the planet to tidal forces, the properties of the materials in the different layers were taken into account through determining the viscosity and rigidity of each layer and weighting them by their volume fractions \citep{Barr2018}. Similarly to Eq. \ref{eq_phi_rho}, the averaged viscosity and shear modulus was approximated as
\begin{align}\label{eq_visc}
            \Phi_{\mathrm{I}}~\eta_{\mathrm{I}} + \Phi_{\mathrm{HPP}}~\eta_{\mathrm{HPP}} + \Phi_{\mathrm{rock}}~\eta_{\mathrm{rock}} \approx \eta, \\
                \Phi_{\mathrm{I}}~\mu_{\mathrm{I}} + \Phi_{\mathrm{HPP}}~\mu_{\mathrm{HPP}} + \Phi_{\mathrm{rock}}~\mu_{\mathrm{rock}} \approx \mu.
        \label{eq_shear}
\end{align}
The planet was then treated as a homogenous body, with the volume-averaged viscosity and rigidity describing its rheological properties. Therefore, despite using only averaged values for $\eta$ and $\mu$ to determine the tidal heat flux in Eq. \ref{eq_tidal}, in reality the effects of the rock, HPP, and ice layers were all taken into account and incorporated in the volume-averaged viscosity and rigidity. Median values of the volume-averaged rheological parameters are shown in the appendix in Table~\ref{tab:app_rheo}. By assuming a uniform heat distribution depending only on the radius of the planet, tidal heat flux could be evaluated at any arbitrary distance from the planetary centre. In order to calculate the temperature of the HPP layer, the heat flux generated by tidal forces was first calculated for the surface using the method described above, then scaled to the rock--HPP boundary with a factor of $(R/R_{\mathrm{rock-HPP}})^2$, where R$_{\mathrm{rock-HPP}}$ is the distance of the boundary layer from the centre of the planet. 

\subsection{Radiogenic heat calculation}

In order to evaluate the interior heat flux, the energy released by the decay of radioactive isotopes with long half-lives was also taken into account. We assumed that radioactive elements are only present in the rock mantle and that these elements are the sole source of radiogenic heating. To calculate the radiogenic heat production rate, we used the formula of \citet{Hussmann2010},
\begin{equation}\label{eq_radheat}
    \dot{E}=m_{\mathrm{sil}}\sum^{4}_{i=1}c_iC_iH_i\mathrm{exp}\lbrack\lambda_i(t-t_{\mathrm{pr}})\rbrack,
\end{equation}
with $i$ indicating the four long-lived isotopes: $^{238}$U, $^{235}$U, $^{232}$Th, and $^{40}$K. The mass of the rock mantle, $m_{\mathrm{sil}} = M_{\mathrm{pl}}\times \mathcal{M_{\mathrm{rock}}}$, where $M_{\mathrm{pl}}$ is the mass of the planet and $\mathcal{M_{\mathrm{rock}}}$ is the mass fraction of the rock mantle,  determined with the interior model (see Section 2.2). The values used for the composition, $c$, and heat release, $H$, of the different elements are shown in Table~\ref{tab:radheat_release} \citep{Audi1997}. Since we did not have information about the abundances ($C$) of these isotopes in the investigated systems, these values are based on different types of meteorites found on Earth \citep{Lodders1998}. We considered two extreme cases to estimate the elemental abundances: CI and LL chondrite abundances. As a lower limit we chose CI chondrite abundances, since this produces the lowest energy rates from the sample. For similar reasons, LL chondrite abundances were chosen as the higher limits. The elemental abundances for each radioactive isotope were then approximated as $C = fC_{\mathrm{CI}} + (1 - f/100)C_{\mathrm{LL}}$, where $f$ was an integer between 0 and 100. The value for $f$ was randomized with uniform probability for each internal structure. In Eq.~\ref{eq_radheat}$, \lambda_i$ is the decay constant, and depends on the half-lives, $\tau_i$, as $\lambda_i = \mathrm{ln~}0.5/\tau_i$. The values of the half-lives and the elemental abundances are shown for all four isotopes in Table \ref{tab:radheat_release}. Since our calculations focused on the current heat production of the exoplanets, here the parameter, $t$, denotes the age of the system, while $t_{\mathrm{pr}}=4.6\times10^9$~Gyr is the age of the Solar System. In the case of GJ~273, K2-72, and Kepler-1649, where the ages of the stars were unknown, $t = t_{\mathrm{pr}}$ was used. The radiogenic heat flux, $F_{\mathrm{rad}}$, was evaluated at the boundary of the rock mantle and the HPP layer using
\begin{equation}
    F_{\mathrm{rad}} = \frac{\dot{E}}{ 4\pi R_{\mathrm{rock-HPP}}}.
\end{equation}

{\renewcommand{\arraystretch}{1.2}
\begin{table}
    \centering
        \caption{Present composition, heat release, half lives, and abundances of the four long-lived radioactive isotopes.}
    \resizebox{\columnwidth}{!}{\begin{tabular}{c c c c c c}
         \hline
         \multirow{2}{*}{Isotope} & \multirow{2}{*}{\textit{c}, kg kg$^{-1}$} & \multirow{2}{*}{\textit{H}, W kg$^{-1}$} & \multirow{2}{*}{$\tau_i$, years} & \multicolumn{2}{c}{C, ppm} \\
          & & & & CI case & LL case\\
         \hline
         \noalign{\smallskip}
         $^{238}$U & 0.992745 & 9.48 $\times$ 10$^{-5}$ & 4.468 $\times$ 10$^{9}$ & 0.008 & 0.015 \\
         $^{235}$U & 7.2 $\times$ 10$^{-3}$ & 5.69 $\times$ 10$^{-4}$ & 0.7038 $\times$ 10$^{9}$ & 0.008 & 0.015 \\
         $^{232}$Th & 1.0 & 2.69 $\times$ 10$^{-5}$ & 15.05 $\times$ 10$^{9}$ & 0.029 & 0.047 \\
         $^{40}$K & 1.17 $\times$ 10$^{-4}$ & 2.92 $\times$ 10$^{-5}$ & 1.277 $\times$ 10$^{9}$ & 550 & 880\\
         \noalign{\smallskip}
         \hline
    \end{tabular}}
    \begin{tablenotes}
        \item[*] Values for $c$, $H$ and $\tau_i$ are from \citet{Audi1997}, while isotope abundances for both CI and LL chondrite cases are from \citet{Lodders1998}.
    \end{tablenotes}
    \label{tab:radheat_release}
\end{table}
}

\subsection{Internal heat dissipation}
We assumed that the heat produced by both tidal interactions and the decay of radioactive isotopes was transported out of the system by solid-state convection. The temperatures at the top and bottom of each convective layer are constant in space due to the 1D nature of our model, but vary as a function of the heat flow across each layer. This accounts for heat transport across the compositional or phase boundary layers by conduction (e.g. \citeauthor{Mueller1988}, \citeyear{Mueller1988}). The convective flux was calculated by using the formula of \citet{Barr2008}: 
\begin{equation}
    F_{\mathrm{conv}} = 0.53k \left( \frac{\rho g\alpha}{\kappa} \right) ^{1/3}\left[\frac{Q}{R_\mathrm{G}T_{\mathrm{hpp}}^2}\right]^{-4/3}\eta_{\mathrm{hpp}}(T_\mathrm{\mathrm{hpp}})^{-1/3},
\end{equation} 
where $k$ is the thermal conductivity, $\rho $ is the density of the HPP layer, $g$ is the local gravitational acceleration, $\alpha$ is the thermal expansion coefficient, $\kappa $ is the thermal diffusivity, $Q $ is the activation energy for ice, $R_\mathrm{G} = 8.314$~J/mol~K is the gas constant, and $T_{\mathrm{hpp}}$ is the temperature of the HPP layer in~K. The viscosity of the HPP layer, $\eta_{\mathrm{hpp}}$, can be expressed as $\eta_{\mathrm{hpp}} = c_{\mathrm{rh}} \exp\left[Q/(R_\mathrm{G}T_{\mathrm{hpp}})\right]$ by following the flow law of \citet{Showman2005}, where $\mathrm{c_{\mathrm{rh}}}=10^{18}/\exp \left[Q/(R_\mathrm{G}T_{\mathrm{melt,hpp}})\right]$ is the rheological constant for the high-pressure ice polymorph, and $T_{\mathrm{melt,hpp}}$ is the melting temperature of the high-pressure ice polymorph, which was assumed to be 280~K in this work. This is a reasonable assumption for the whole layer that spans from the bottom of the ice I layer at 209~MPa to pressures around 2.5 GPa, where the Ice VII phase dominates and above which the rheology of ice is not well constrained. In reality, the melting temperature of the HPP is the lowest at the top of the HPP layer (253~K) and the highest at the lower boundary. The values of the physical parameters of the HPP layer are shown in Table~\ref{tab:ice_params}. 

The equilibrium temperature of the HPP layer, T$_{\mathrm{hpp}}$, was calculated by equating the convective flux with the internal heat flux at the rock--HPP boundary: 
\begin{equation}
    F_{\mathrm{conv}} = F_{\mathrm{int}} = F_{\mathrm{tid}} +  F_{\mathrm{rad}},
\end{equation}
where both tidal and radiogenic heat fluxes represent values scaled by the area of the rock-HPP boundary surface (see section 2.3.). We assumed that H$_2$O was present in liquid form underneath the surface layer, when the equilibrium temperature of the HPP layer was above the melting point of the high-pressure ice polymorph. 

\begin{table}
    \centering
       \caption{Physical parameters of the high-pressure ice polymorph (HPP) layer.}
    \setlength\tabcolsep{2pt}
    \begin{tabular}{c c c}
    \hline
         & HPP layer & References \\
         \hline
         \noalign{\smallskip}
         Thermal conductivity, \textit{k} [W/m/K] & 651/T & 1,2 \\
         Thermal expansion coefficient, $\alpha$ [1/K] &  $1.6 \times 10^{-4}$ & 2,3 \\
         Thermal diffusivity, $\kappa$ [m$^2/$s] & $10^{-6}$ & 2,4 \\
         Activation energy, \textit{Q} [kJ/mol] & 20 &\\
         Melting temperature, $T_{\mathrm{melt,hpp}}$ [K] & 280 & 5 \\
         Density, $\rho$ [kg/m$^3$] & 1300 & 3, 6\\
    \noalign{\smallskip}
         \hline     
    \end{tabular}
    \begin{tablenotes}
        \item[*] References: 1: \citet{Slack1980}, 2: \citet{Kirk1987}, 3: \citet{Hobbs1974}, 4: \citet{Peddinti2019}, 5: \citet{Durham1997}, 6: \citet{Barr2018}
    \end{tablenotes}
    \label{tab:ice_params}
\end{table}

\subsection{Summary of the analysis}

Here we present a brief summary of how we applied the methods detailed in the previous sections. We modelled the interiors of 28 HZ exoplanets based on their measured parameters. Planetary masses and radii were randomly generated with the Forecaster model using an empirical relation, if either one was unknown. Since a large number of interior structures were possible for a given density, the generated mass-radius pairs corresponded to tens to hundreds of thousands of distinct possible interiors per exoplanet.

We calculated the total internal heat flux at the rock-HPP boundary for each internal composition. We assumed two different heat sources: tidal heating and radiogenic heating. For the latter, elemental abundances were approximated using values measured in CI and LL chondrite meteorites. We note that other heat sources, such as secular cooling of the core or greenhouse warming above the surface may be additional contributors to the heating of the interior. Our calculations therefore serve as lower estimates for the total internal heat flux. We calculated the temperature of the HPP layer by assuming that the internal heat flux was transported out of the system via solid-state convection. If the equilibrium temperature of the HPP layer was above the melting point of the HPP ice, we concluded that the HPP layer was in liquid form. This was calculated for all the possible interior structures of each planet.

\section{Results}

\subsection{H$_2$O content of the planets}
We estimated the H$_2$O content of all 28 planets based on their modelled interiors. For a given internal structure, the H$_2$O mass fraction is defined as the sum of the mass fractions of the surface ice/water layer and the underlying HPP layer. Since a large number of different internal structures were feasible for a given planet, our calculations resulted in a wide range of H$_2$O mass fractions for each planet. 

In Figure \ref{fig:h2o_mfrac} the distribution of the resulting internal structures in different H$_2$O mass fraction ranges for each planet is shown. Each box represents a 3$\%$ mass fraction interval in the top panel of the figure and 6$\%$ in the bottom panel, respectively. The colour and the numbers indicate the fraction of all modelled interiors that resulted in internal structures within a certain H$_2$O mass fraction range. If a large number of interiors had high H$_2$O mass fractions for a given planet, that appears as a shift in the values and lighter colours in our figure towards the higher mass fraction ranges. If the input parameters - the mass and radius of the planet - had low uncertainties, the resulting H$_2$O mass fractions concentrate in a relatively narrow range, with the H$_2$O mass fractions showing a peak in the distribution at the largest probabilities. For planets with less constrained parameters, the H$_2$O mass fractions spanned a broader range of feasible values (see for example K2-3~d, K2-72~e or Kepler-1229~b). 

\begin{figure}[hbt!]
    \centering
    \includegraphics[width=0.8\columnwidth]{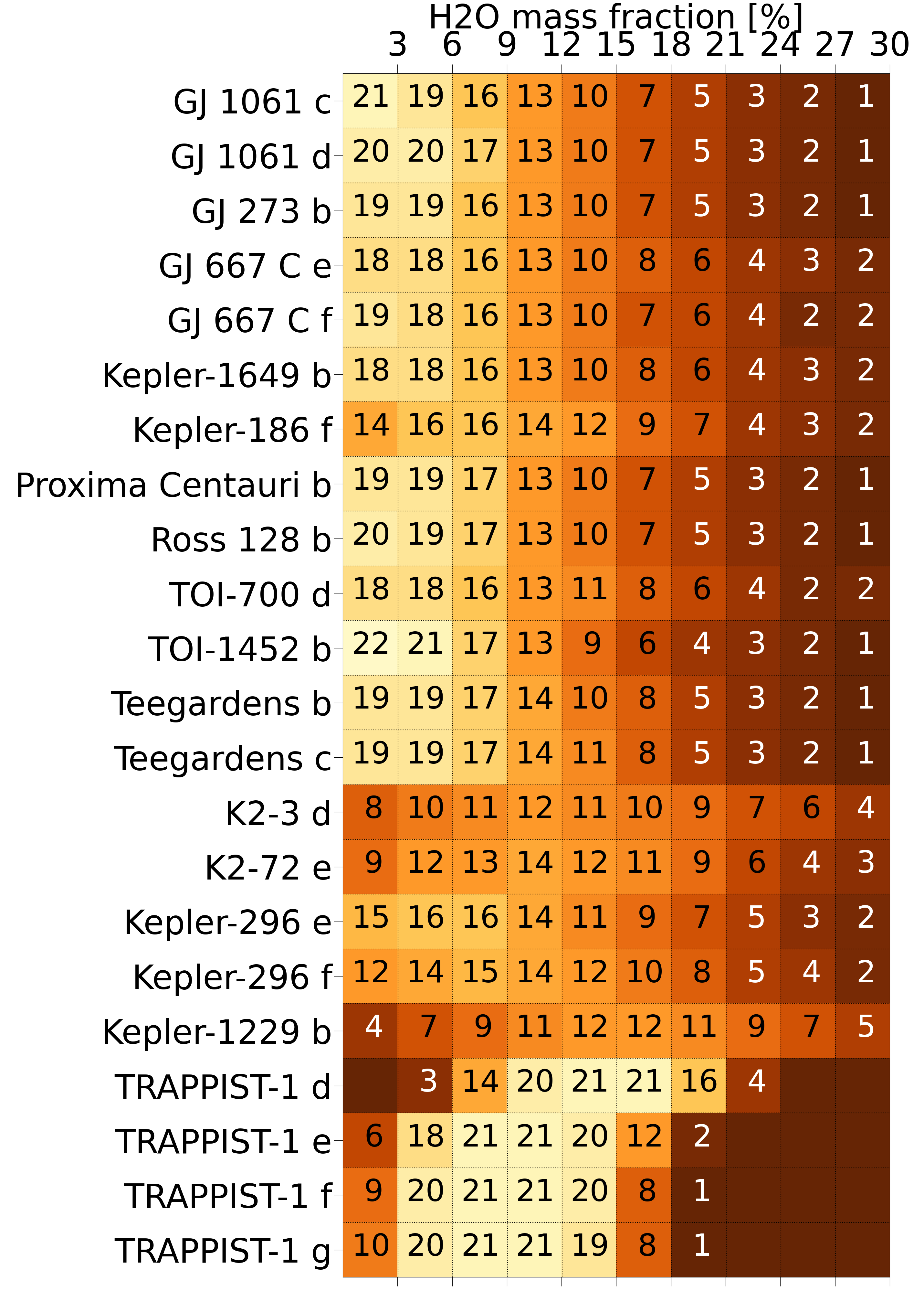}
    \includegraphics[width=0.8\columnwidth]{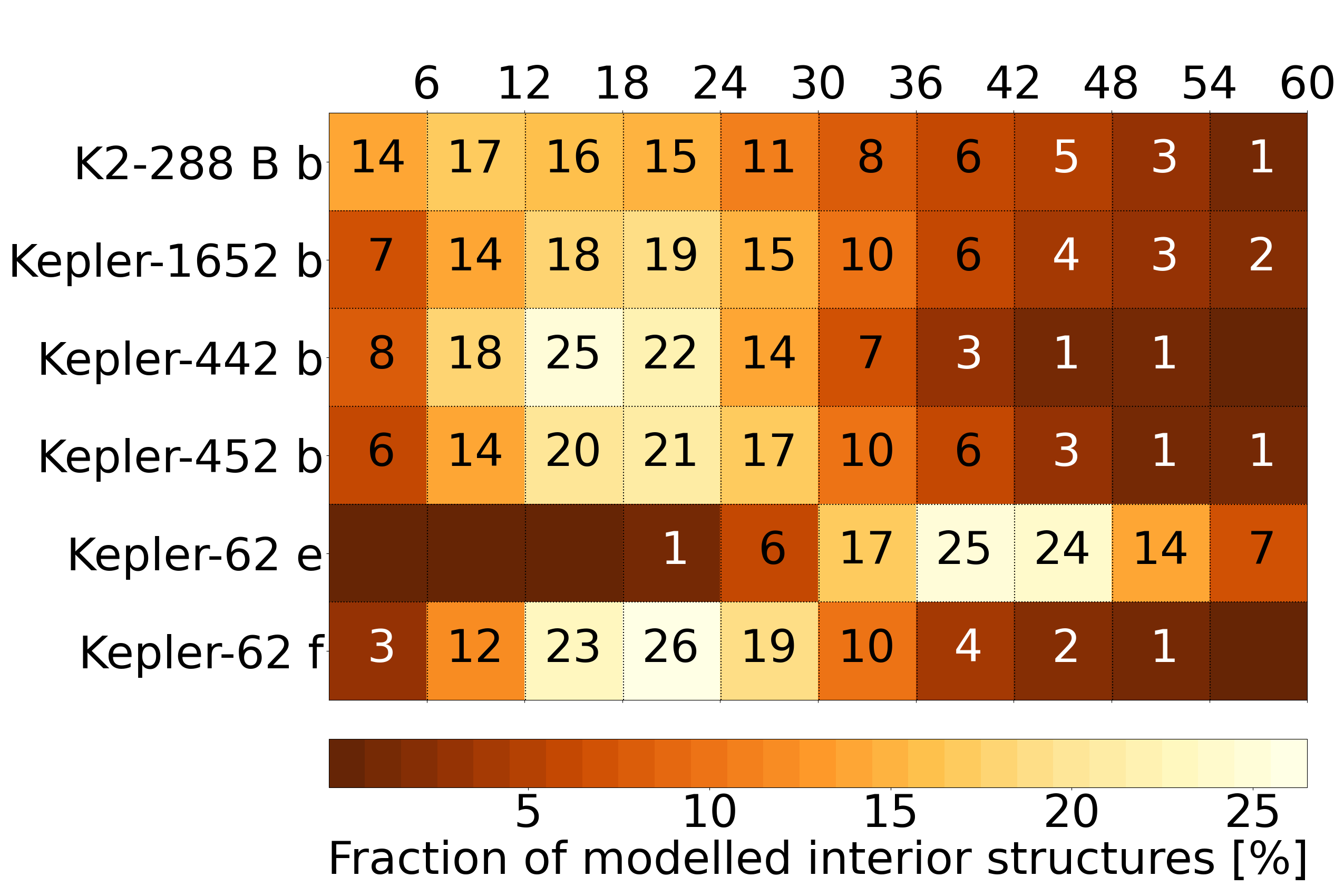}
    \caption{Distributions of the modelled H$_2$O contents of the investigated exoplanets. Here the H$_2$O mass fractions are divided into 3$\%$ or 6$\%$ groups in the top and bottom panels, respectively. The colour indicates the fraction of internal structures within the given percentage groups of the water mass fraction, with the values written on the figure. Probabilities of less than 1\% are not shown. Planets with water mass fractions over 30$\%$ are shown in the bottom panel. We note the different scales in the two panels.}
    \label{fig:h2o_mfrac}
\end{figure}

Our results show that all of these planets could have large enough H$_2$O mass fractions to have global ice/water surfaces. Planets with surface temperatures above the melting point of ice and large water mass fractions (e.g. Kepler-296~e) may have global oceans regardless of the presence of an atmosphere. The equilibrium temperatures (T$_\mathrm{eq}$) of the investigated systems are shown in Table~\ref{tab:measurements}. It is important to note that the method for estimating T$_\mathrm{eq}$ may differ for the individual planets in the assumed heat distribution and Bond albedos. For more details, the reader is referred to the papers referenced in Table~\ref{tab:measurements}. For those planets where the surface temperature falls below the melting point of ice, liquid water may not be present on the surface in the absence of an atmosphere, and large water mass fractions may imply global ice layers. If the internal heat flux in these bodies is sufficiently high to cause melting in the HPP layer, these worlds may harbour underground liquid water reservoirs. Interactions between the rock and water layer may then result in a suitable environment for life to develop, which makes these planets interesting targets for habitability studies. Since the investigated planets all lie within the HZ, high values for H$_2$O mass fractions could also imply global oceans on the surface if the planets have atmospheres similar to that of the Earth.

For some planets, like GJ 1061 c and d, the thickness of the ice layer could not be properly determined, since our model resulted in a similar number of cases in a wide range of water mass fractions. This arose from the fact that the planetary parameters of these bodies were not known to sufficiently high precision. In order to refine our results for these planets, more precise measurements of the planetary masses and radii are required.

For planets with well-constrained planetary masses and radii, such as the TRAPPIST-1 system, the interior structures were constrained to a higher degree, which results in more precise H$_2$O-content estimations. All of the modelled TRAPPIST-1 planets are likely to have extended H$_2$O layers. Since TRAPPIST-1 e, f, g, and potentially planet d as well are in the HZ of their star, determining the state of the H$_2$O layer is necessary to assess the habitability of these planets.

In some cases the modelled exoplanets had extremely large H$_2$O mass fractions, which may imply that those planets fall under the category of ocean worlds. Although our model did not account for planetary atmospheres, HZ exoplanets with such high water mass fractions are likely to have surface oceans, since water vapour will provide non-negligible atmospheric pressure. Our results suggest that Kepler-62~f, Kepler-452~b and Kepler-442~b could be members of the ocean world class (bottom panel of Figure \ref{fig:h2o_mfrac}). In the case of Kepler-62~e and K2-288~B~b, most of the generated mass-radius pairs result in mini-Neptunian compositions. However, if indeed superterrans, our models show that these planets may also be part of the ocean world planetary group. For each planet, the probabilities of having rocky compositions (here referred to as \textit{terran probability}) are shown in the seventh column of Table \ref{tab:summary}.

\subsection{Effects of tidal and radiogenic heating}

The separate effects of radiogenic and tidal heating at the rock-HPP ice boundary are shown in Figure~\ref{fig:tid_rad} and in Table~\ref{tab:summary}. The median values of radiogenic and tidal heat fluxes are shown in Table~\ref{tab:summary} along with values corresponding to the first and third quartiles. Boxplots for tidal and radiogenic heating are shown in Figure~\ref{fig:tid_rad}. The top panel depicts planets with strong tidal heating, while the bottom panel shows exoplanets where radiogenic heat flux is comparable with or exceeds the flux from tidal heating. The maximum values correspond to $\mathrm{Q3} + 1.5\times\mathrm{IQR}$, while the minima are defined as $\mathrm{Q1} - 1.5\times\mathrm{IQR}$, where $\mathrm{IQR} = \mathrm{Q3} - \mathrm{Q1}$ is the interquartile range. In the case of radiogenic heating, elemental abundances were in every case a mix of CI and LL chondritic abundances. Tidal heating, if it was present in the planet, was almost always at least an order of magnitude higher than radiogenic heating and was the main source of internal heat for close-in exoplanets around late-type stars, as seen in the top panel of Figure \ref{fig:tid_rad}. Tidal heat flux on these planets, like GJ 1061 c and d, was close to 2~W$/\mathrm{m}^2$, similar to the heat flux observed on Io \citep{Veeder2004}. In some planets, as in the case of Kepler-296~e, the average heat flux from the two different sources is comparable.

On average, the internal heat flux provided by radiogenic heating was significantly lower compared to the tidal heat flux in close-in exoplanets. Melting of the HPP ice layer, however, still took place even in the absence of tidal interactions on many of the investigated exoplanets, as seen in Table~\ref{tab:summary}. In the case of some exoplanets with radiogenic heating as the main internal heat source, the assumed abundances for radioactive elements may be the decisive factor for melting to occur. More data on radioactive elemental abundances in exoplanetary systems may improve our understanding of the interior heating in these bodies. 

\begin{figure}
    \centering
    \includegraphics[width=\columnwidth,height=8.2cm, keepaspectratio]{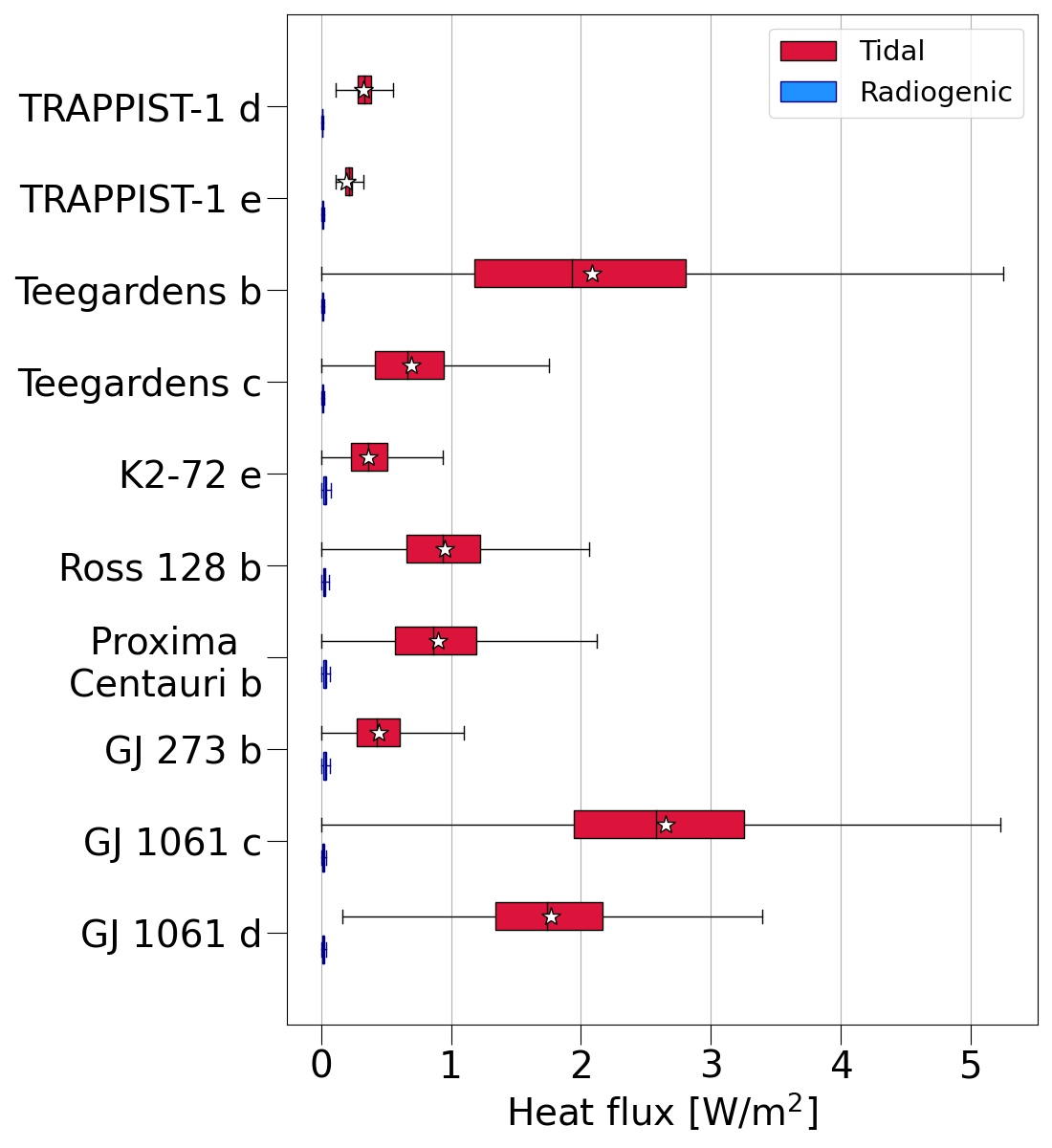}
    \includegraphics[width=\columnwidth,height=12.4cm, keepaspectratio]{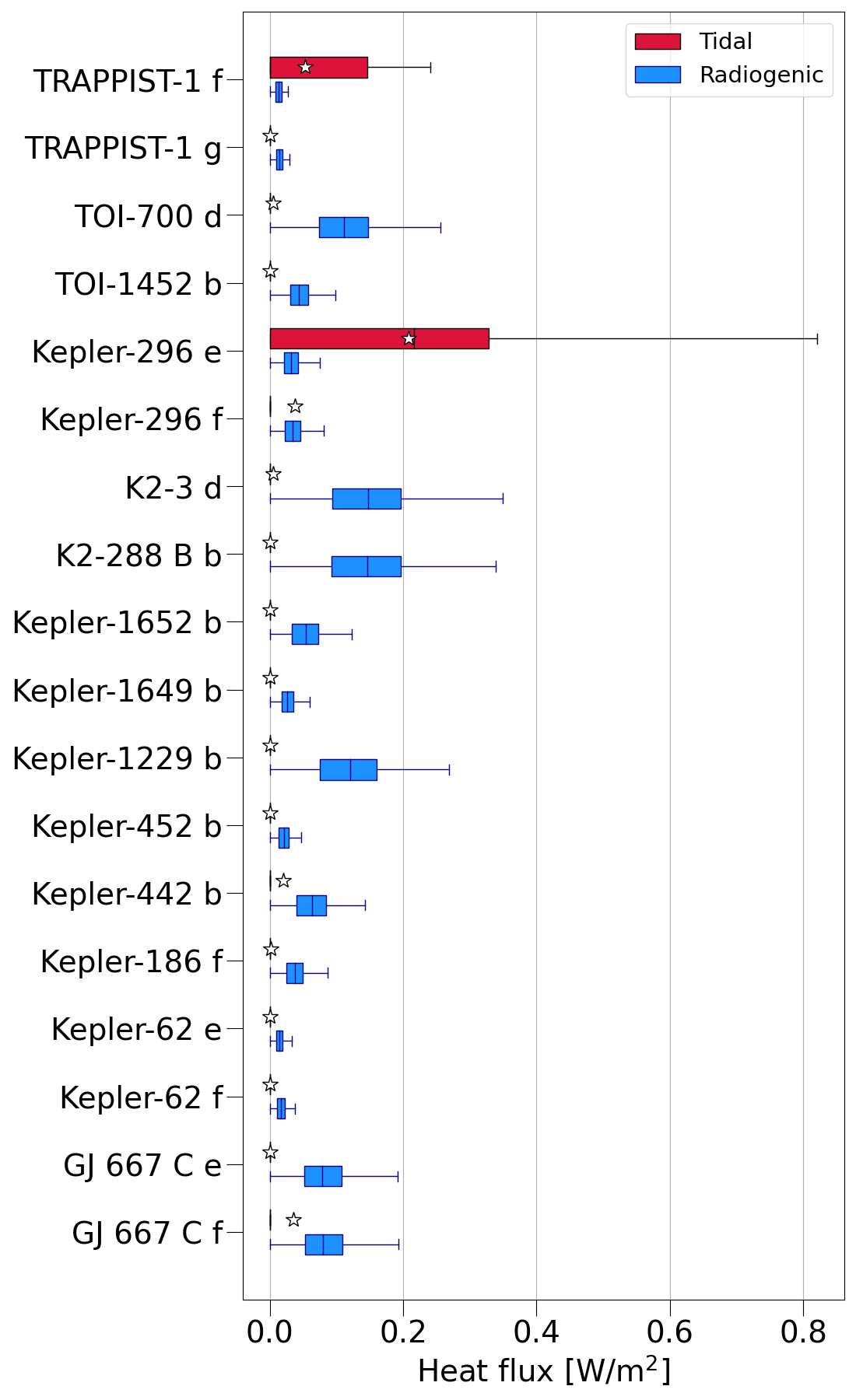}
    \caption{Contribution of the different heat sources to the internal heat flux at the rock--HPP boundary for each planet. The red boxes indicate tidal heating, while blue denotes radiogenic heat flux. Median values are represented as vertical lines inside the boxes. The upper and lower boundaries of the boxes correspond to the third and first quartiles (Q3 and Q1), respectively, while error bars show the maximum and minimum values. White stars indicate the average tidal heat flux values for every planet. We note the different scales in the two panels.}
    \label{fig:tid_rad}
\end{figure}

Planets with considerable tidal heating (shown in the top panel of Figure~\ref{fig:tid_rad}) were further investigated, in order to have a better insight into the possible tidal heating rates. For these planets the distribution of the tidal heat flux at the rock-HPP boundary is shown in Figure~\ref{fig:tidal_hist}. Tidal heat flux values shown here were averaged for all mass-radius pairs in the case of every planet, which gave a total of maximum 10000 different tidal heat flux values. Structures where tidal heating had no contribution to the internal heat budget were possible for all exoplanets, though to different degrees for each body (represented by the height of the peaks at zero W/m$^2$). The shape of the distribution, as well as the probable heat flux values, also differed for the investigated planets. It is clearly seen from the histograms that the maximum values for the tidal heating rates of many of these planets (e.g. GJ~1061~c and d) were not representative values, but were in fact outliers in the distribution.

\subsection{Melting caused by internal heating}

The phase of H$_2$O in the HPP layer was further investigated. If the combined internal heat flux from tidal heating and the decay of radioactive nuclei for a given internal structure was high enough to heat up the HPP layer above the melting point, it resulted in a water layer below the surface ice. Here, the melting probability is the fraction of all modelled structures that resulted in T$_{\mathrm{HPP}}>$T$_{\mathrm{melt,HPP}}$. Melting probabilities for each planet are shown in Table~\ref{tab:summary}. The higher the melting probability for a certain planet, the more likely that the HPP layer is melted.

The effects of tidal and radiogenic heating on the melting of the HPP layer were studied in further detail. For close-in HZ exoplanets, for example GJ 1061 c and TRAPPIST-1 d, tidal heating was efficient enough to melt the underlying ice layer in most of the cases. This caused a higher overall fraction of modelled structures to experience melting than in the case of exoplanets farther out from their host stars, where radiogenic heating was the main contributor to the internal heat budget. For planets in the HZs of more massive stars, tidal dissipation is not a substantial source of energy, since for these planets the probability of being captured into spin-orbit resonance in small.

Since the radiogenic heat production rate depends on the thickness of the rock mantle, different internal structures resulted in different internal heat fluxes. Differences in the internal structures therefore affected the resulting radiogenic heat flux. Further uncertainties in the melting probabilities originated from the unknown abundances of radioactive elements in the investigated exoplanetary systems. In this study these abundances were approximated based on a mixture of elemental abundances seen in Solar System meteorites, with the two extremes being the purely LL chondritic abundances (providing the maximum possible heat flux for exoplanets dominated by radiogenic heating) and the purely CI chondritic abundances (associated with the lowest heating rates, see section 2.4).

\subsection{Overall likelihood of containing liquid water}

In order to properly assess the potential of a water reservoir under the surface of these planets, it is important to take into account other factors, such as the likelihood of the planet having a rocky composition. The values for terran probability, in other words the probability of a planet having a rocky composition, are represented by the fraction of generated planetary masses below 2.04 $M_\oplus$ for each planet. This boundary mass is based on the transition point between rocky and Neptune-like planets in the Forecaster model. On the other hand, in a recent study \citet{Luque2022} argue that for planets around M dwarf stars, the density of the exoplanet is the key parameter that separates rocky planets, water worlds, and mini-Neptunes from each other, not the planetary mass or radius. However, for the sake of consistency, in this present study the boundary mass based on the transition point in the Forecaster code was used for classification in all cases. The only exception was TOI-1452~b, where both the planetary mass and radius with symmetric errors were known. In this case, even though the mass of the planet was over the 2.04 $M_\oplus$ transition point, the density suggested a terrestrial composition in all cases. Since neither the mass nor the radius had to be approximated using the Forecaster code, this did not result in any inconsistency. It is important to note that while some planets, for example Kepler-442 b, may have an extended H$_2$O-layer that is likely to melt due to internal heating processes, the planet itself has a low probability of being a rocky planet. It is evident that the best targets for habitability studies are those exoplanets that are highly likely to have rocky composition, have a high chance of having an H$_2$O-layer, and experience moderate internal heat flux. 

To identify the most likely candidates for having a liquid water layer, we defined the ocean probability parameter ($P_{\mathrm{ocean}}$) as
\begin{equation}
    P_{\mathrm{ocean}} = P_{\mathrm{terran}} * P_{\mathrm{melt}},
\end{equation}
 where $P_{\mathrm{terran}}$ is the terran probability and $P_{\mathrm{melt}}$ is the melting probability defined in Section~3.3. The $P_{\mathrm{ocean}}$ parameters are shown in Table~\ref{tab:summary} for each of the investigated planets. Our calculations show that Ross~128~b, Teegarden's~b, Proxima~Centauri~b, TRAPPIST-1~d, GJ~1061~c and~d, Teegarden's~c, and TRAPPIST-1~e have the largest values for $P_{\mathrm{ocean}}$, and hence are the most likely to have extended internal water reservoirs.

Figure~\ref{fig:melt_percent} shows $P_{\mathrm{ocean}}$ for the investigated systems, along with the HZ for stars with effective temperatures above 2750~K. The HZ is shown for illustrative purposes only. The boundaries of the HZ were calculated based on the stellar parameters (mass, temperature, and luminosity) of 2.5 Gyr old stars from the MESA Isochrones and Stellar Tracks (MIST) online tool \citep{Dotter2016, Choi2016, Paxton2018}. A second-order polynomial was fitted to the data using the method of \citet{Kopparapu2014} for planets with masses of 1~$M_{\oplus}$, which provided the boundary of the HZ for stars above 2600~K. The HZ of TRAPPIST-1 is not shown in the figure because the temperature of the star is below this value, and this method could not be used to estimate its HZ.  


\section{Discussion and conclusions}

\begin{figure*}
    \centering
    \begin{subfigure}{0.19\textwidth}
        \centering
        \includegraphics[width=\textwidth]{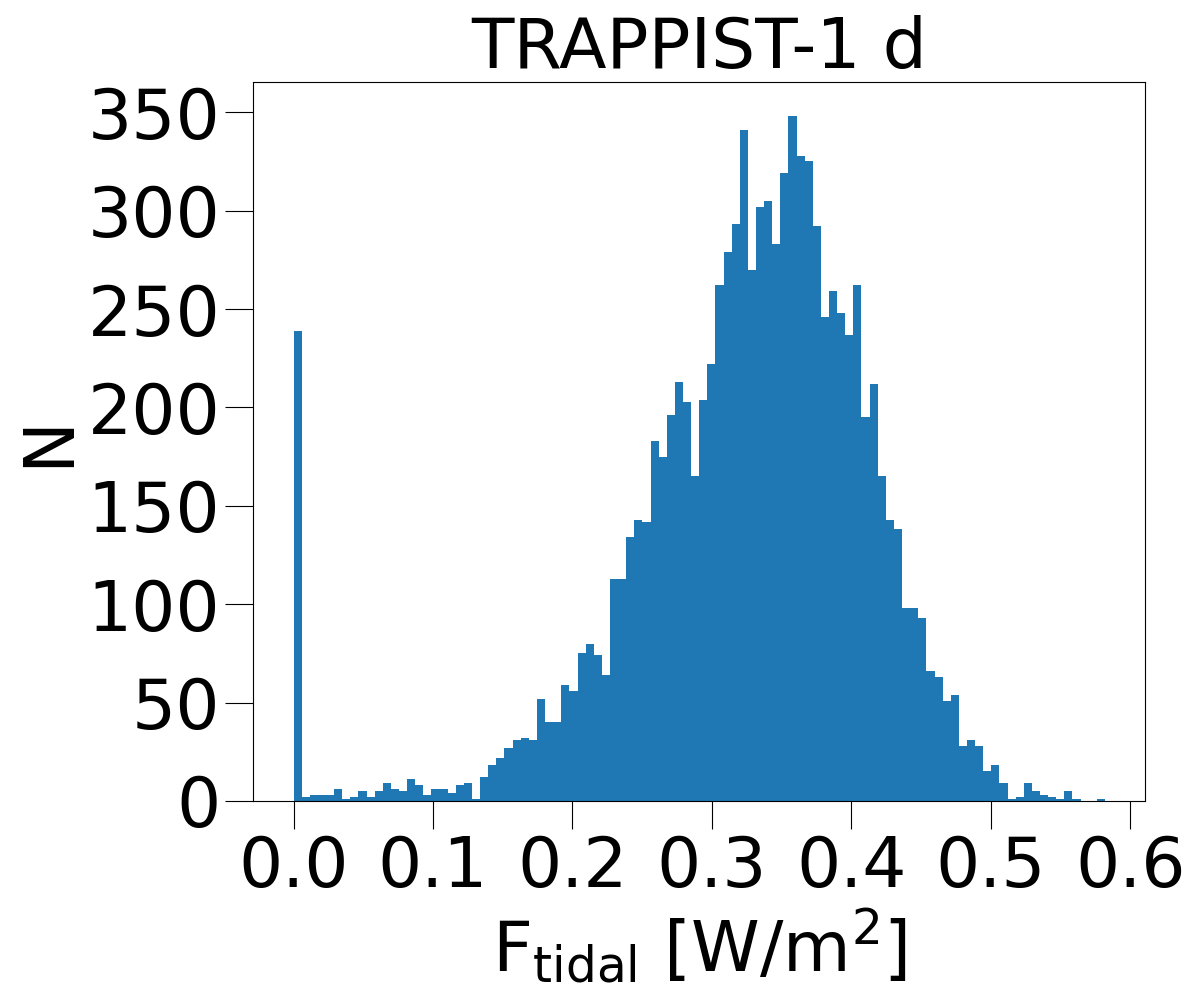}
    \end{subfigure}
    \begin{subfigure}{0.19\textwidth}
        \centering
        \includegraphics[width=\textwidth]{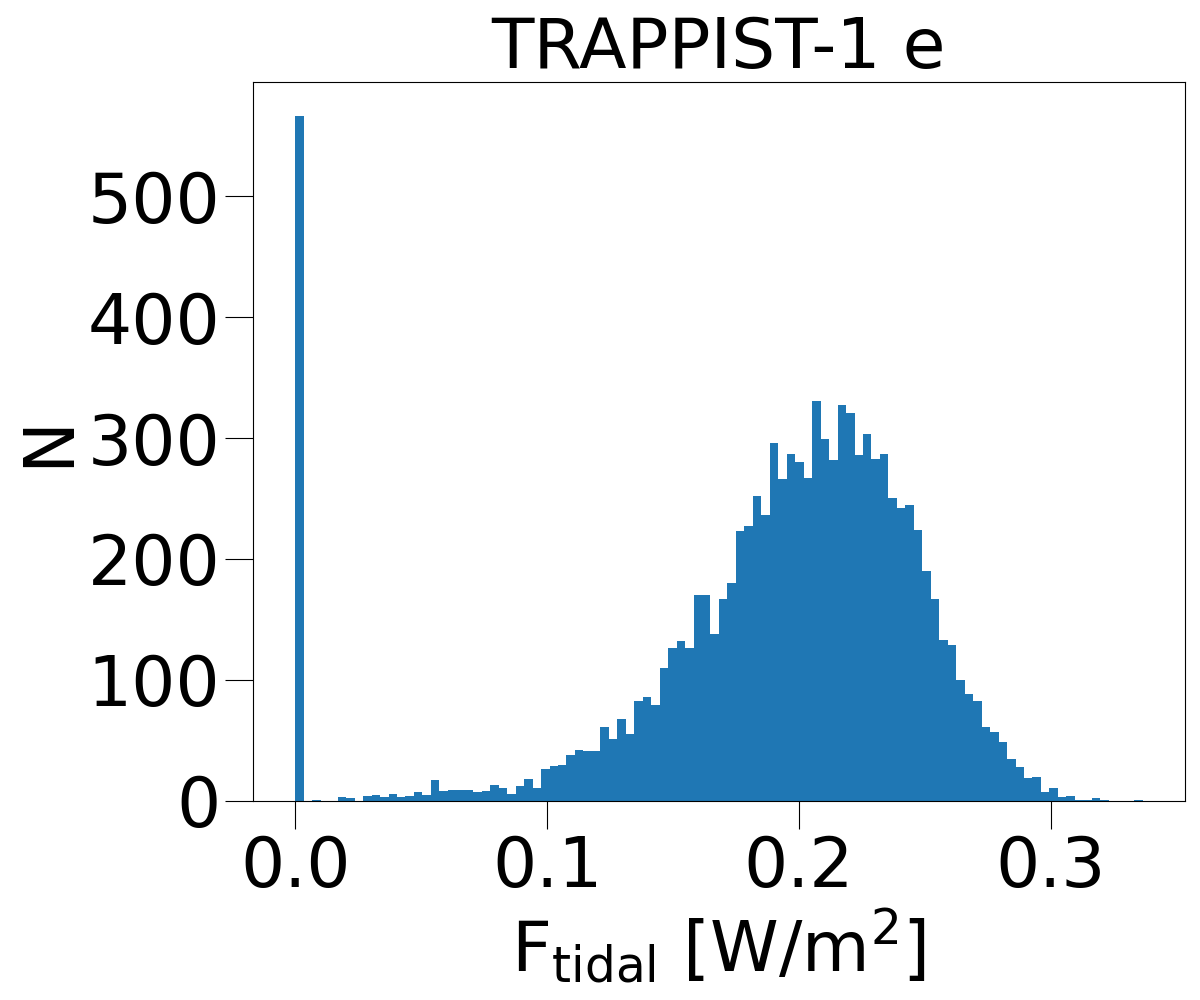}
    \end{subfigure}
    \begin{subfigure}{0.19\textwidth}
        \centering
        \includegraphics[width=\textwidth]{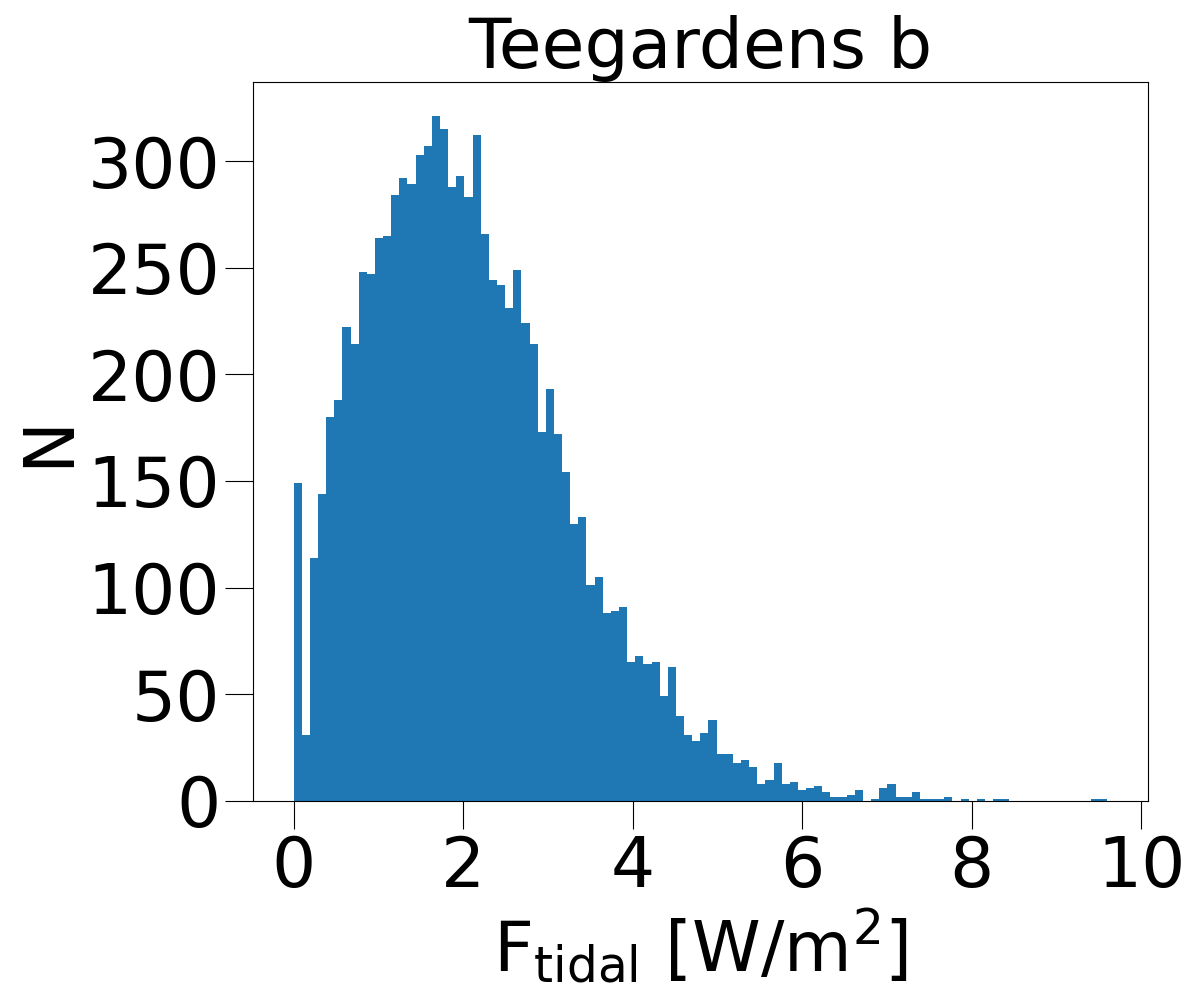}
    \end{subfigure}
    \begin{subfigure}{0.19\textwidth}
        \centering
        \includegraphics[width=\textwidth]{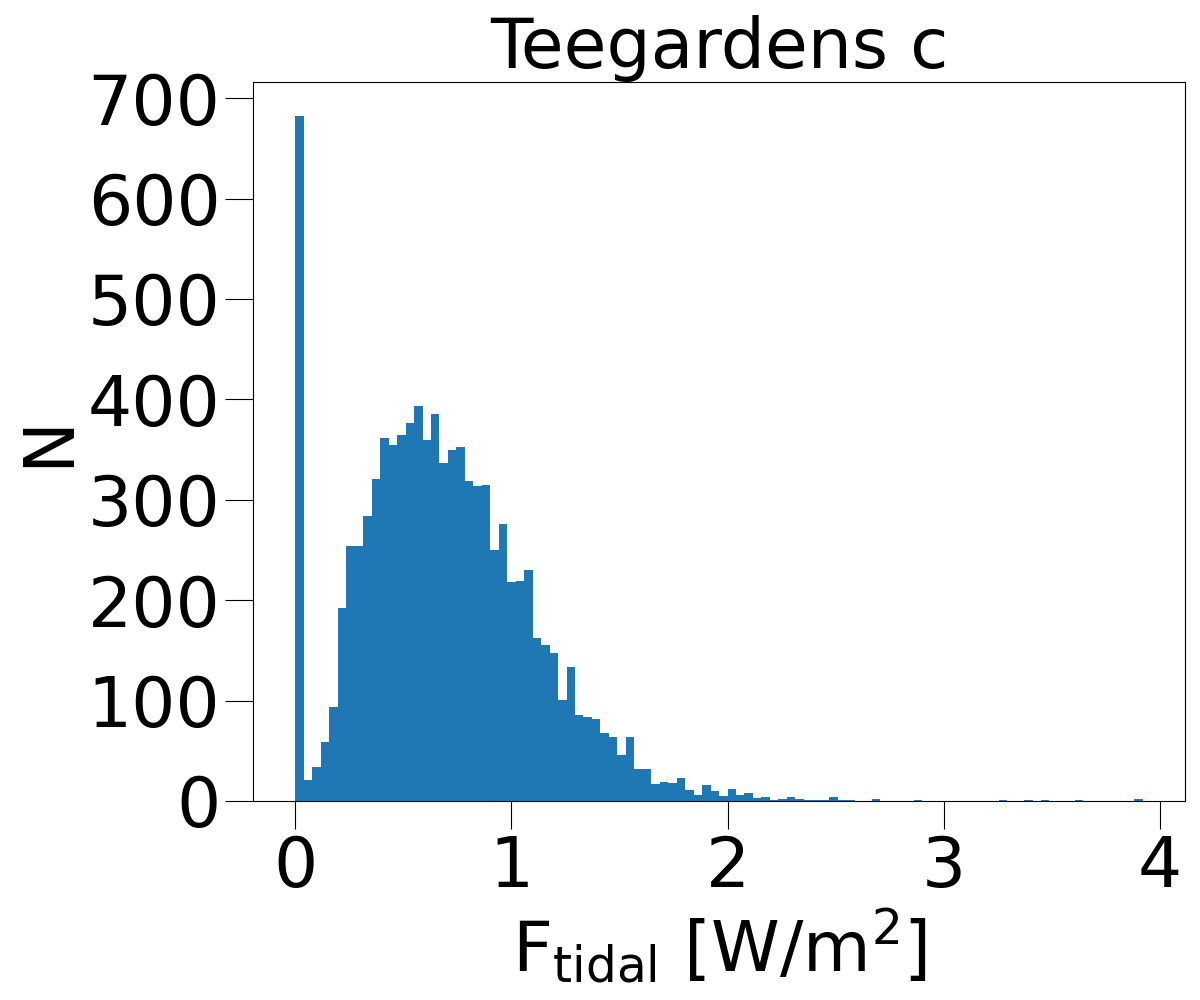}
    \end{subfigure}
    \begin{subfigure}{0.19\textwidth}
        \centering
        \includegraphics[width=\textwidth]{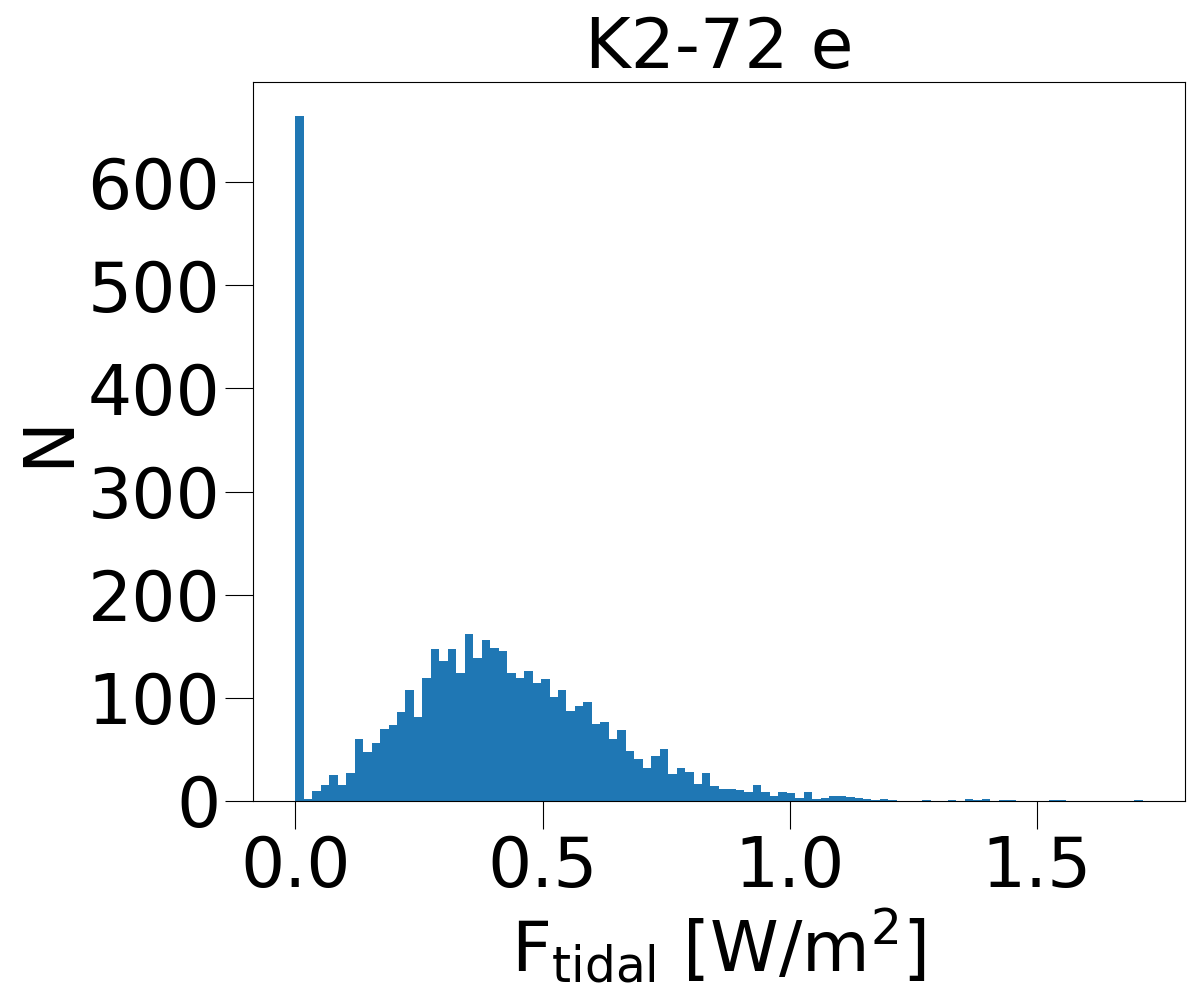}
    \end{subfigure}
    \begin{subfigure}{0.19\textwidth}
        \centering
        \includegraphics[width=\textwidth]{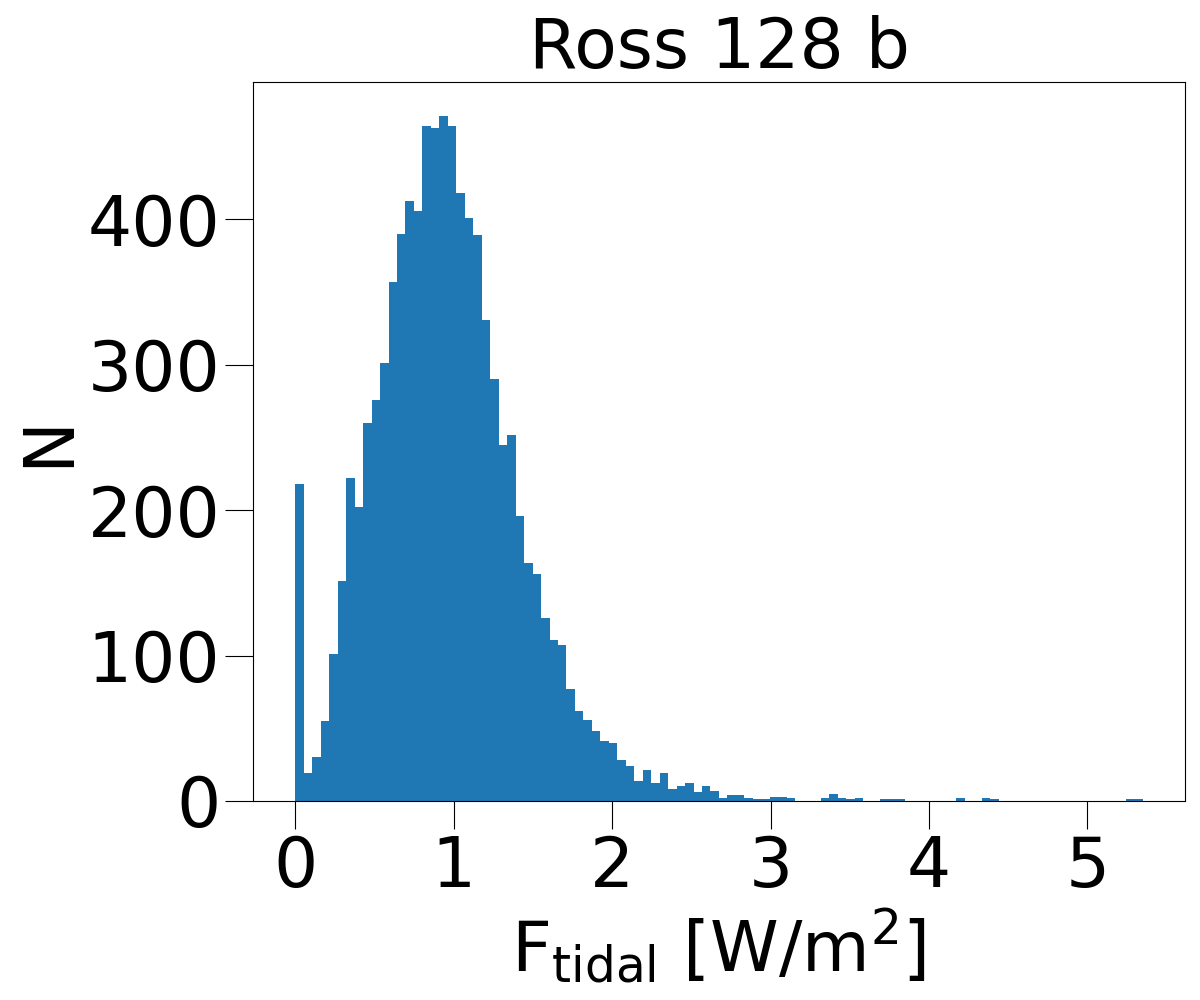}
    \end{subfigure}
    \begin{subfigure}{0.19\textwidth}
        \centering
        \includegraphics[width=\textwidth]{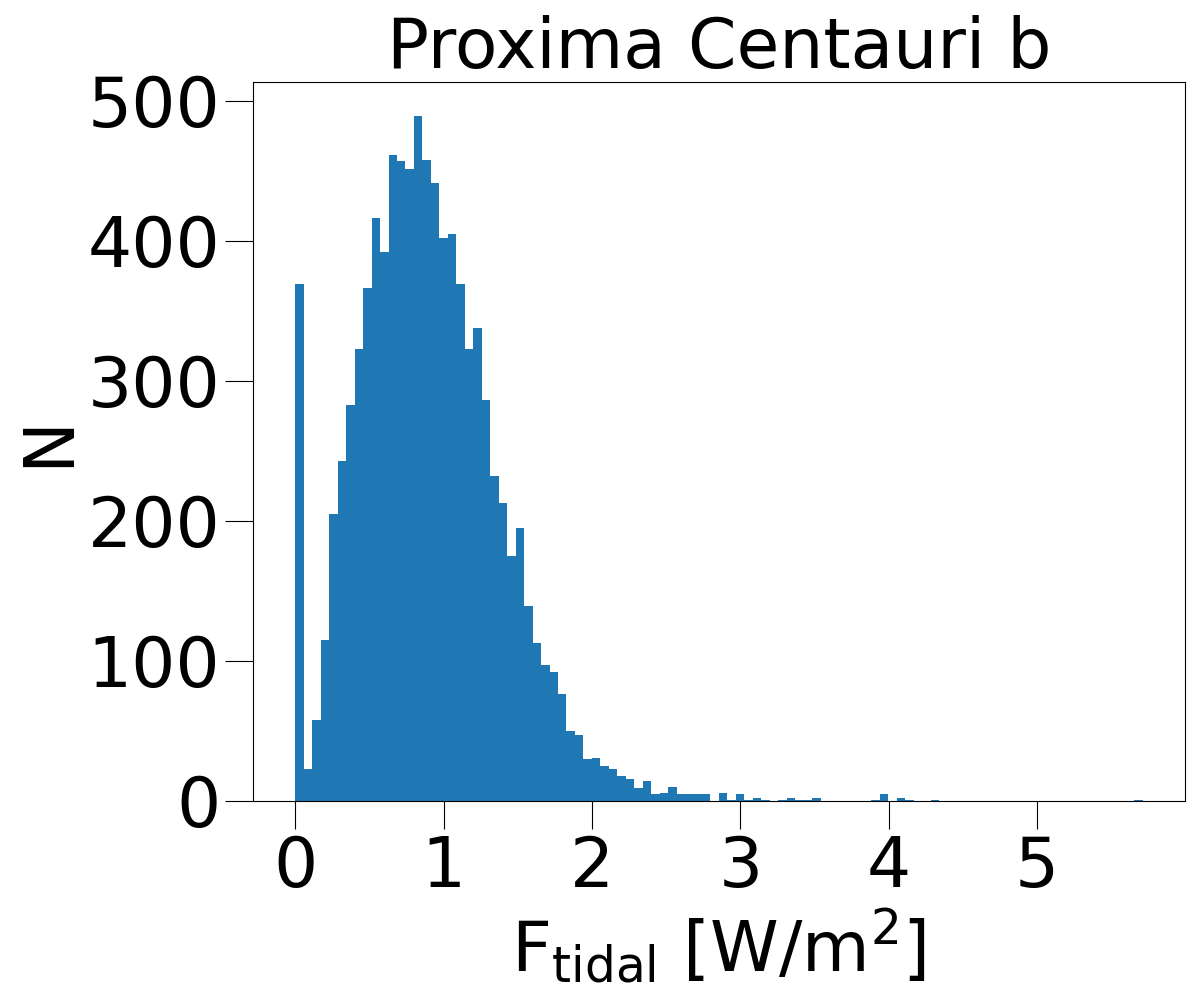}
    \end{subfigure}
    \begin{subfigure}{0.19\textwidth}
        \centering
        \includegraphics[width=\textwidth]{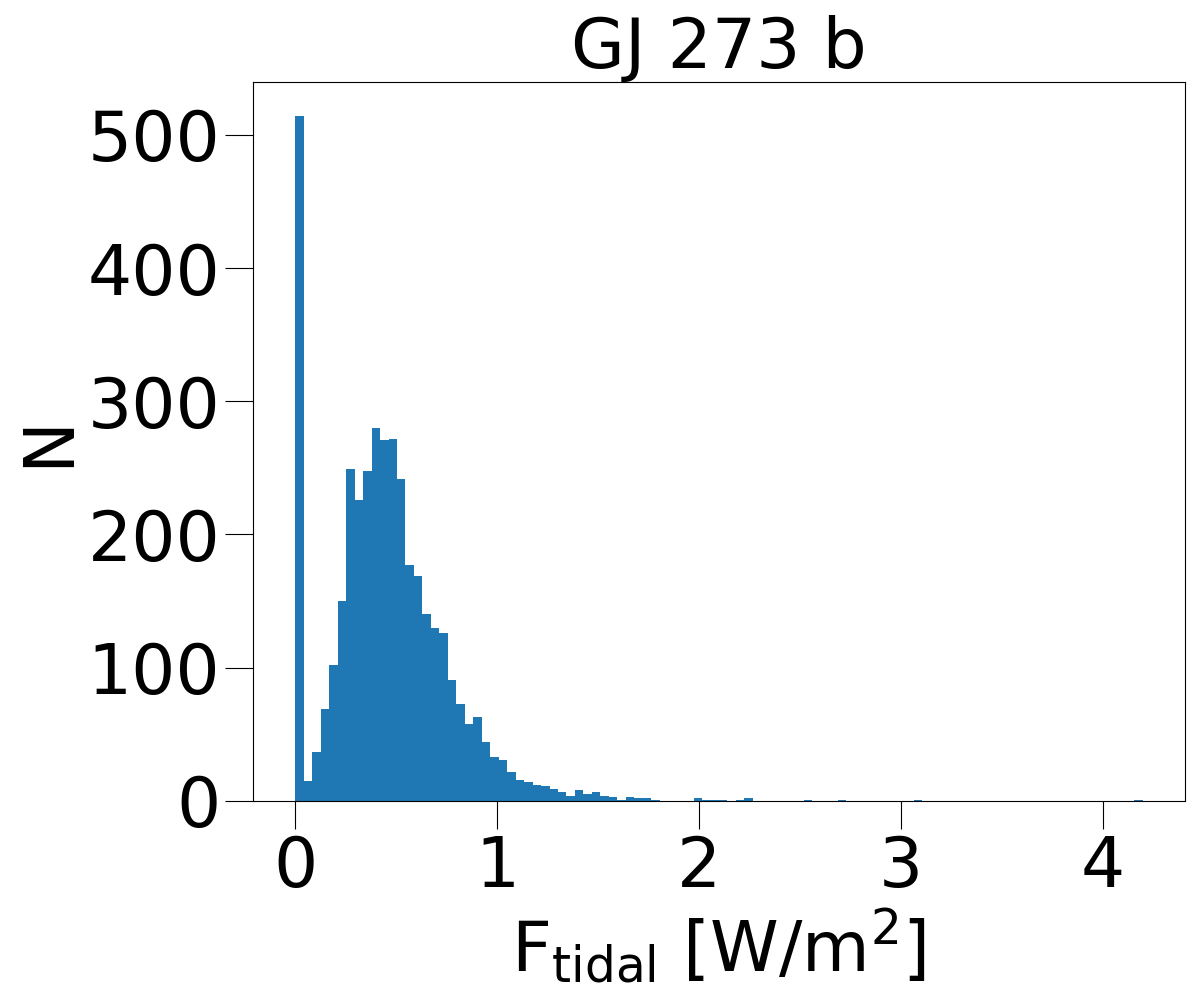}
    \end{subfigure}
    \begin{subfigure}{0.19\textwidth}
        \centering
        \includegraphics[width=\textwidth]{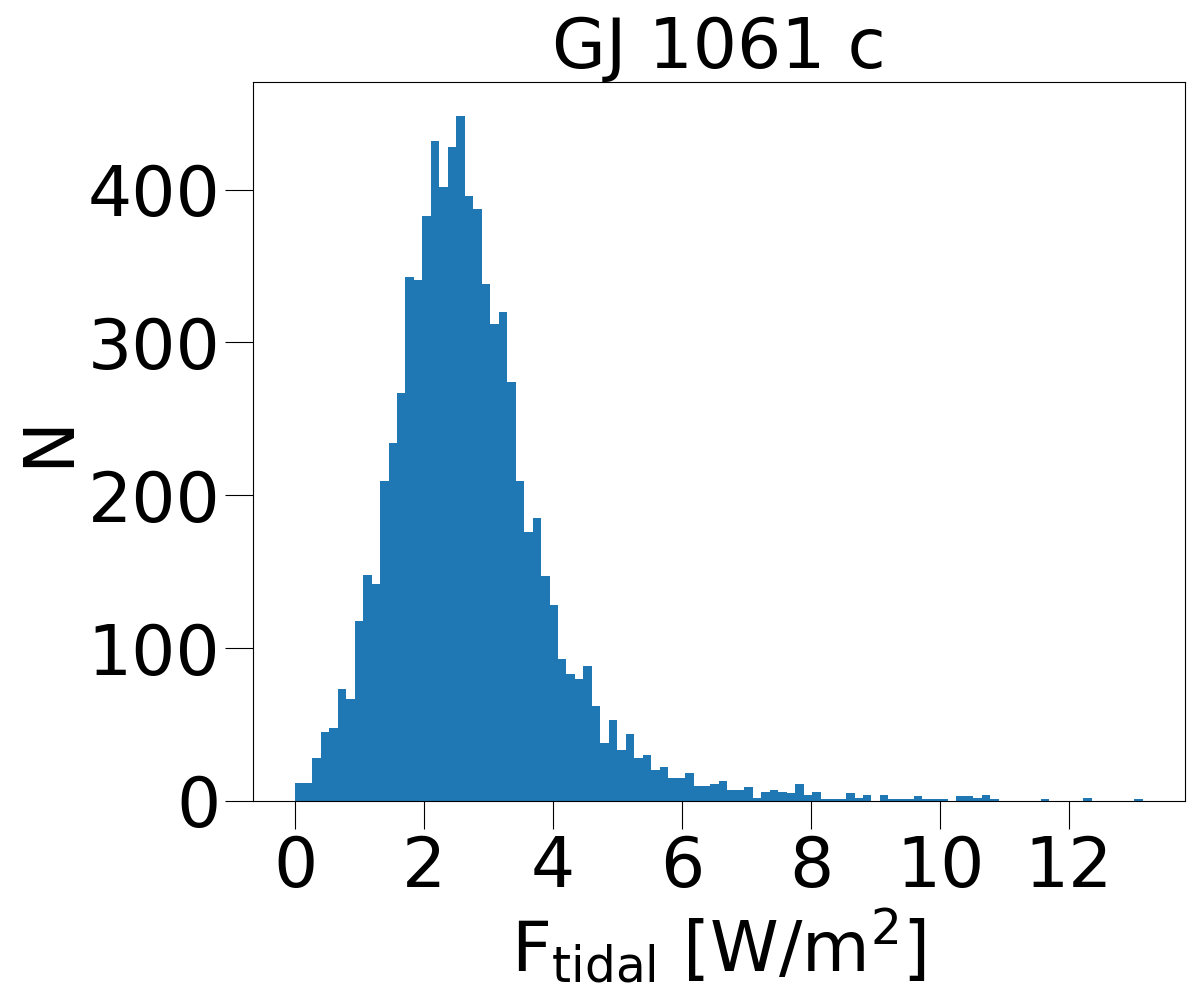}
    \end{subfigure}
    \begin{subfigure}{0.19\textwidth}
        \centering
        \includegraphics[width=\textwidth]{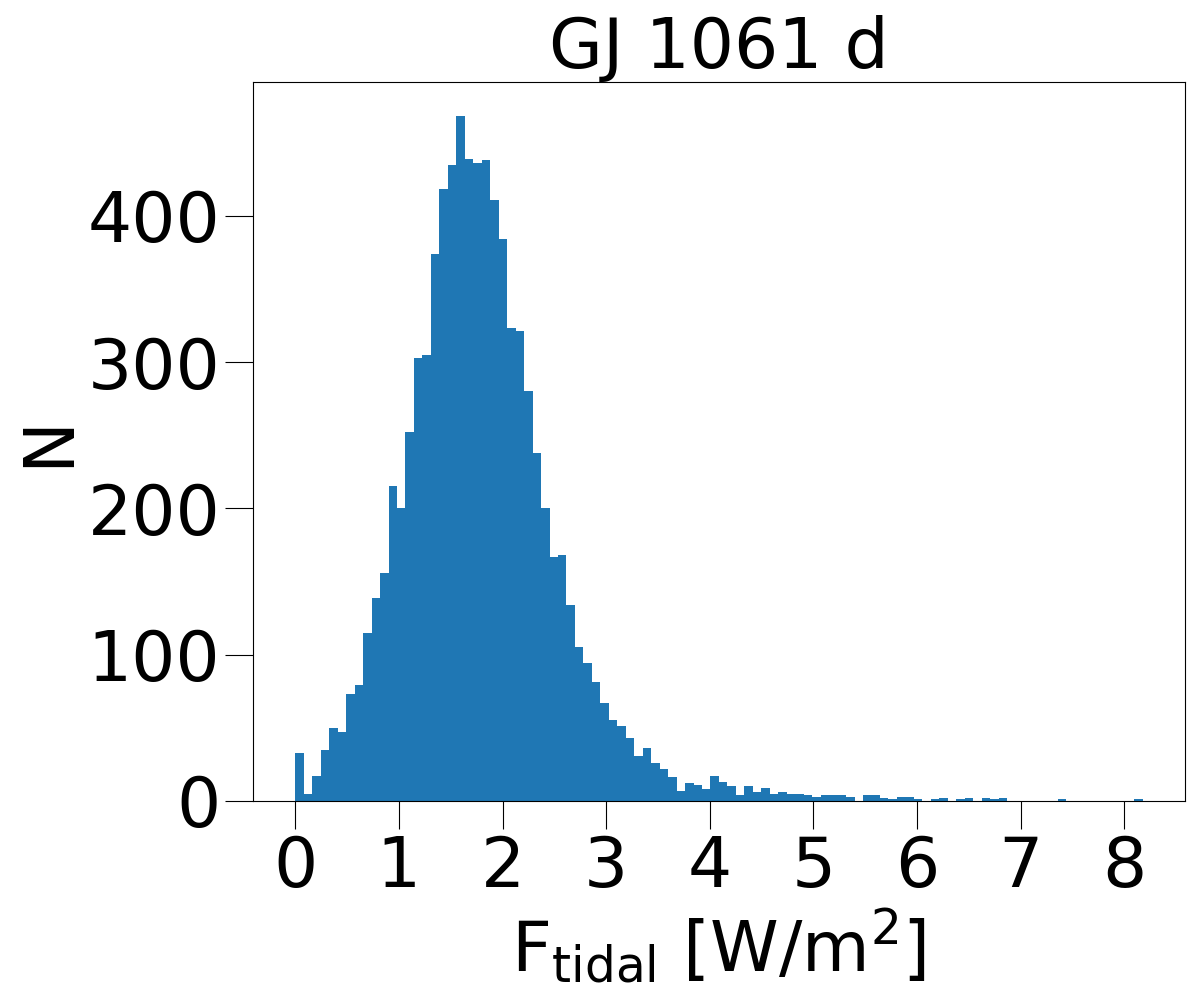}
    \end{subfigure}
    \caption{Histograms showing the distribution of the tidal heat flux values, F$_{\mathrm{tidal}}$, calculated at the rock-HPP boundary for planets with strong tidal heating, with N denoting number density.}
    \label{fig:tidal_hist}
\end{figure*}

\begin{figure*}
    \centering
    \includegraphics[width=\textwidth]{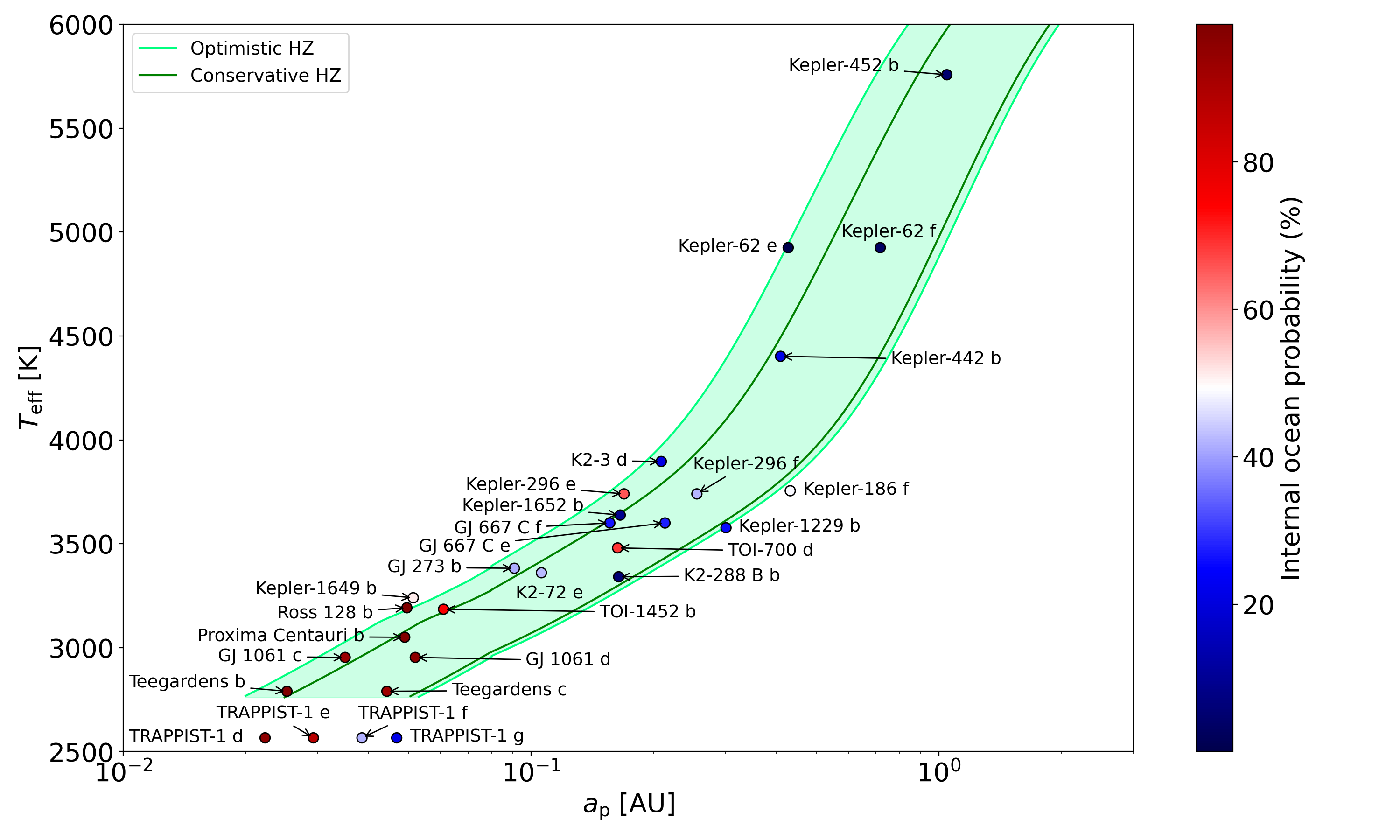}
    \caption{Internal ocean probabilities of the investigated exoplanets. The colour bar depicts the probability of each planet having a liquid water reservoir in its interior. Red colours imply a high probability, while planets with dark blue circles are less likely to have internal oceans. The HZ, shown as a green band only for illustrative purposes, was calculated based on stellar parameters from theoretical stellar evolution models.}
    \label{fig:melt_percent}
\end{figure*}

Our research provides an insight into the conditions of the interiors of 28 HZ rocky exoplanets from the viewpoint of habitability. We modelled the interior structure of each planet and determined the possible H$_2$O content of these bodies. Internal heat fluxes from the combined effect of tidal and radiogenic heating were calculated for all exoplanets. We estimated the temperature of the HPP ice layer based on the internal heat flux and calculated the probability of the HPP layer melting. Our calculations suggest that planets with the highest probabilities of containing liquid water reservoirs are Teegarden's~b and c, GJ~1061~c and d, TRAPPIST-1~d and e, Ross~128~b, and Proxima~Centauri~b. These exoplanets are likely to be rocky planets while having an extended water layer, and hence are of outstanding importance to habitability studies.

\begin{table*}
    \centering
    {\renewcommand{\arraystretch}{1.4}
        \caption{Internal heat fluxes, melting, and terran and ocean probabilities of planets.}
            \label{tab:summary}
    \begin{tabular}{c c c c c c }
         \hline 
         \multirow{2}{*}{Planet} & \multirow{2}{*}{F$_{\mathrm{tid}}$, W/m$^2$} & \multirow{2}{*}{F$_{\mathrm{rad}}$, W/m$^2$}  &  \multirow{2}{*}{$P_{\mathrm{melt}}$, $\%$} & \multirow{2}{*}{$P_{\mathrm{terran}}$, $\%$} & \multirow{2}{*}{$P_{\mathrm{ocean}}$, $\%$} \\
         & & & & & \\
         \hline 
         \noalign{\smallskip}
         Ross 128 b & 0.94$_{0.66}^{1.22}$ & 0.03$_{0.02}^{0.03}$ & 98.71 & 99.93 & 98.65\\ 
         Teegardens b & 1.93$_{1.18}^{2.81}$ & 0.01$_{0.01}^{0.01}$ & 98.45 & 100.00 & 98.45\\
         Proxima Centauri b & 0.87$_{0.57}^{1.19}$ & 0.03$_{0.02}^{0.04}$ & 98.16 & 100.00 & 98.16\\ 
         GJ 1061 d & 1.74$_{1.34}^{2.16}$ & 0.02$_{0.01}^{0.02}$ & 99.65 & 96.61 & 96.27\\
         TRAPPIST-1 d & 0.34$_{0.28}^{0.39}$ & 0.01$_{0.01}^{0.01}$ & 95.91 & 100.00 & 95.91\\ 
         GJ 1061 c & 2.58$_{1.95}^{3.26}$ & 0.02$_{0.01}^{0.02}$ & 99.88 & 93.88 & 93.77\\
         Teegardens c & 0.66$_{0.41}^{0.95}$ & 0.01$_{0.01}^{0.01}$ & 92.84 & 100.00 & 92.84\\
         TRAPPIST-1 e & 0.21$_{0.18}^{0.24}$ & 0.01$_{0.01}^{0.02}$ & 87.56 & 100.00 & 87.56\\
         TOI-1452 b & 0.0$_{0.0}^{0.0}$ & 0.04$_{0.03}^{0.06}$ & 75.03 & 100.00 & 75.03\\ 
         TOI-700 d & 0.0$_{0.0}^{0.0}$ & 0.11$_{0.07}^{0.15}$ & 93.52 & 73.67 & 68.89\\
         Kepler-296 e & 0.22$_{0.00}^{0.33}$ & 0.03$_{0.02}^{0.04}$ & 84.96 & 77.41 & 65.77\\
         Kepler-1649 b & 0.0$_{0.0}^{0.0}$ & 0.03$_{0.02}^{0.03}$ & 58.18 & 88.08 & 51.25\\
         Kepler-186 f & 0.0$_{0.0}^{0.0}$ & 0.04$_{0.02}^{0.05}$ & 70.35 & 69.34 & 48.78\\ 
         K2-72 e & 0.36$_{0.23}^{0.51}$ & 0.03$_{0.02}^{0.04}$ & 91.67 & 46.59 & 42.71\\
         Kepler-296 f & 0.0$_{0.0}^{0.0}$ & 0.03$_{0.02}^{0.05}$ & 70.11 & 60.03 & 42.09\\
         TRAPPIST-1 f & 0.0$_{0.0}^{0.15}$ & 0.01$_{0.01}^{0.02}$ & 41.90 & 100.00 & 41.90\\
         GJ 273 b & 0.43$_{0.28}^{0.61}$ & 0.03$_{0.02}^{0.04}$ & 93.79 & 44.00 & 41.27\\
         GJ 667 C e & 0.0$_{0.0}^{0.0}$ & 0.08$_{0.05}^{0.11}$ & 90.39 & 30.42 & 27.50\\    
         GJ 667 C f & 0.0$_{0.0}^{0.0}$ & 0.08$_{0.05}^{0.11}$ & 91.61 & 29.43 & 26.96\\
         Kepler-1229 b & 0.0$_{0.0}^{0.0}$ & 0.12$_{0.08}^{0.16}$ & 89.51 & 28.49 & 25.50\\
         TRAPPIST-1 g & 0.0$_{0.0}^{0.0}$ & 0.01$_{0.01}^{0.02}$ & 21.88 & 100.00 & 21.88\\
         Kepler-442 b & 0.0$_{0.0}^{0.0}$ & 0.06$_{0.04}^{0.08}$ & 77.61 & 27.49 & 21.34\\
         K2-3 d & 0.0$_{0.0}^{0.0}$ & 0.15$_{0.09}^{0.20}$ & 92.94 & 22.66 & 21.06\\
         Kepler-1652 b & 0.0$_{0.0}^{0.0}$ & 0.05$_{0.03}^{0.07}$ & 68.64 & 13.00 & 8.92\\
         K2-288 B b & 0.0$_{0.0}^{0.0}$ & 0.15$_{0.09}^{0.20}$ & 90.41 & 6.30 & 5.70\\
         Kepler-452 b & 0.0$_{0.0}^{0.0}$ & 0.02$_{0.01}^{0.03}$ & 27.21 & 17.23 & 4.69\\
         Kepler-62 f & 0.0$_{0.0}^{0.0}$ & 0.02$_{0.01}^{0.02}$ & 14.31 & 24.01 & 3.44\\ 
         Kepler-62 e & 0.0$_{0.0}^{0.0}$ & 0.01$_{0.01}^{0.02}$ & 0.28 & 7.82 & 0.02\\ 
         \noalign{\smallskip}
         \hline
    \end{tabular}}
    \begin{tablenotes}
        \item[] Note: Median values for tidal and radiogenic heat flux at the rock-HPP boundary are summarized for each planet. The first and third quartiles are show in the lower and upper index of the median value, respectively. Melting probabilities ($P_{\mathrm{melt}}$) are shown in the fourth column. The $P_{\mathrm{terran}}$ column shows the probability of each planet having a rocky composition based on its measured and generated masses and radii. The values for the ocean probabilities ($P_{\mathrm{ocean}}$) are shown in the last column.
    \end{tablenotes}
\end{table*}

An interesting remark is that all these planets orbit M dwarf stars. This, however, does not mean that exoplanets of M dwarfs are more likely to have subsurface oceans in general, but rather it implies that there is some kind of an underlying bias in our sample. A possible explanation is that our current exoplanet detection methods are more sensitive to finding planets with lower masses or of smaller sizes orbiting closer to their host stars than rocky worlds of similar sizes further out in their systems. Hence, HZ exoplanets with smaller sizes or masses are more likely to be detected around M dwarf stars. Since $P_{\mathrm{terran}}$ is determined based on planetary masses, low-mass bodies are expected to have larger probabilities of having rocky compositions than planets with higher measured sizes or masses. This could explain why planets in our sample that reside in the HZ of earlier-type (F,G,K) stars have smaller $P_{\mathrm{ocean}}$ - a parameter that is directly proportional with $P_{\mathrm{terran}}$ - than exoplanets orbiting M dwarf stars.

There were a number of exoplanets in our sample (such as GJ~1061~c and d) where the internal heating processes proved to be strong enough to induce melting in more than 98\% of all modelled interiors. These planets are of particular interest, since melting is almost guaranteed to take place. With the currently available mass and radius measurements, however, it is not possible to properly assess the probability of a thick H$_2$O layer being present on these planets. Although the results of our modelling permitted the presence of an extended ice layer, the H$_2$O mass fraction of the planet could not be adequately constrained. To have a greater insight into the structures of these potentially habitable bodies, more precise observations of their masses and radii are required. 

In order to properly estimate the radiogenic heat flux, a better understanding of the average elemental composition in the investigated systems is required. A more realistic approximation of the radioactive isotope abundances may improve our knowledge of the internal heat production in these exoplanets. In accordance with previous studies on this subject, which showed that solar analogues may be more abundant in radioactive elements than our Sun, we assume that our calculations do not overestimate the radiogenic heat flux in the investigated systems. 

It is worth noting that tidal heating is strongly dependent on the eccentricity of the planetary orbit, which changes during the evolution of the planet. As a result of tidal forces, the orbit of close-in exoplanets may eventually become circular ($e=0$). Tidal heating may therefore completely diminish without any external perturbation that would help maintain non-zero eccentricities. However,  $e>0$ may prevail due to the gravitational effects of other planets in the system. For example, in the TRAPPIST-1 system, mean motion resonances prevented the circulization of planetary orbits. The eccentricities of the TRAPPIST-1 planets are still above zero, despite the old age of the system.

Since the atmosphere contributes little to the mean density of rocky planets compared to the mantle or the iron core \citep{Noack2017}, for our approach of modelling the interiors of exoplanets the presence of an atmosphere was not assumed. The knowledge of whether the planet has an extended atmosphere or not, however, could help in the interpretation of our results. One such planet is Kepler-62~f, where an extended, low-density atmosphere could explain the lower mean density of the planet, which in our modelling was interpreted as extremely high water content. Furthermore, liquid oceans on the surface of rocky planets with low water mass fractions can be present if the planets have atmospheres. On the other hand, planets with high water mass fractions and equilibrium surface temperatures above the melting point of ice may have global oceans on their surfaces even without an extended atmosphere and regardless of internal heating. TRAPPIST-1~d and Kepler-296~e may be such bodies, where the evaporation of a possible surface ocean itself may be sufficient to produce a low-pressure water vapour atmosphere. However, planets without a surface ice layer, and with internal heat flux high enough to induce a runaway greenhouse state, could end up losing their water \citep{Barnes2013}. The effect of tidal heating on the habitability of exoplanets and its possible impact on the boundaries of the HZ is further explored in \citep{Barnes2009}.

Modelling the interior of these rocky bodies gives us a more accurate picture of the structure and internal conditions of HZ exoplanets. Determining the H$_2$O content and the internal heating of these planets allows us to estimate the probability of a large water reservoir and opens a new window for habitability studies. Subsurface oceans may be excellent sites for the development of life, although this may be difficult to observe with our current technology. 

According to \citet{Tian2015}, planets with Earth-like water content around M dwarf stars may be rare because of water loss due to the high XUV radiation of the star. However, they found that ocean planets
with a water mass fraction higher than 10\% are able to retain their
water, a fact that coincides with our results on the water-rich
interiors of the TRAPPIST-1 planets. Recent studies argue that the origin of high water content in planets around low-mass stars can be explained by migration models \citep{Unterborn2018, Miguel2020}. \citet{Unterborn2018} found that the TRAPPIST-1 planets may have formed outside of the snow line and then migrated inward. On the other hand, water transport through asteroid bombardment cannot be ruled out as well. Bombardment in compact systems, which could be responsible for a smaller fraction of the total water content observed in these planets, was found to be possible through various methods, for example assuming a giant planet within the system or a perturbing body outside of the stellar system  \citep{Dosovic2020, Clement2022}. 

\citet{Brugger2016} modelled the structure of Proxima Centauri b assuming five different layers - a metallic core, a lower and an upper mantle, a HPP layer, and a surface H$_2$O layer. They concluded that even internal structures with water mass fractions of 50\% are possible for Proxima Centauri b. \citet{Herath2021} investigated the interior structures of Ross~128~b and Proxima~Centauri~b and - in agreement with \citet{Brugger2016} for the latter and with our results for both exoplanets - showed that high volatile content in these planets cannot be excluded. They investigated scenarios with different water mass fractions in the case of both planets, and concluded that a liquid water layer can be present over a thick ice shell for structures with water mass fractions larger than 5\%, assuming a planetary atmosphere. Our results also suggest that there is a high chance that Proxima~Centauri~b may have a liquid ocean layer, although, contrary to \citet{Brugger2016}, in our study this liquid water layer is beneath the ice shell. Nevertheless, in the presence of an atmosphere, our results are consistent with a surface ocean layer. Furthermore, \citet{Brugger2016} also find that high water content can hinder the generation of a magnetic field in these planets. Notably, they conclude that water mass fractions larger than 10\% make the presence of liquid outer cores unlikely. 

Interior modelling for TOI-1452~b by \citet{Cadieux2022} suggests that the planet may be a water world with a high water mass fraction of $22_{-13}^{+21}\%$. In our study, however, scenarios with a significantly lower water mass fraction are favoured, although high values like those presented by \citet{Cadieux2022} cannot be excluded either.

In agreement with our study, \citet{Jackson2008hab} found that tidal heating may be beneficial for the subsurface habitability of icy bodies by allowing the existence of an ocean under the surface ice layer. 

Besides modelling the interior of HZ exoplanets, our approach is also suitable for the characterization of rocky exoplanets farther outside the HZs of their stars. However, the only “cold” rocky exoplanets known to date are those discovered by gravitational microlensing. Once such exoplanets with the potential of further characterization are discovered, our method could be applied to gain more insight into the internal conditions of those bodies.

\begin{acknowledgements}
The authors thank the help of Zoltán Dencs, Michaela Walterová, Eike Guenther and René Heller for discussing the results of the manuscript.

\'AB and VD have been supported by the Hungarian National Research, Development, and Innovation Office (NKFIH) grant K-131508. The COFUND project oLife has received funding from the European Union's Horizon 2020 research and innovation programme under grant agreement No 847675.

This research has made use of the Planet Habitability Laboratory website hosted by the University of Puerto Rico at Arecibo.
\end{acknowledgements}

%
%

\bibliography{references.bib}

\begin{appendix}
\section{Volume-averaged rheological parameters}
    \begin{table}[h]
    \renewcommand{\arraystretch}{1.5}
    \centering
        \caption{Volume-averaged rheological parameters of exoplanets with non-zero tidal heating.}
        \setlength\tabcolsep{1.9pt}
    \begin{tabular}{c c c}
        \hline
        Name & $\mu$, GPa & $\eta$, $10^{12}$ Pa s \\
        \hline
        \noalign{\smallskip}
        GJ 1061 c & 0.02$_{0.01}^{0.04}$ & 30$_{14}^{73}$ \\
        GJ 1061 d & 0.05$_{0.02}^{0.10}$ & 88$_{41}^{200}$ \\
        GJ 273 b & 1.10$_{0.48}^{2.68}$ & 3148$_{1280}^{8287}$ \\
        K2-72 e & 1.43$_{0.62}^{3.54}$ & 4255$_{1722}^{11288}$ \\
        Kepler-296 e & 3.13$_{1.41}^{6.87}$ & 9948$_{4244}^{22633}$ \\
        Proxima Centauri b & 0.21$_{0.09}^{0.57}$ & 503$_{201}^{1527}$ \\
        Ross 128 b & 0.20$_{0.10}^{0.49}$ & 477$_{208}^{1273}$ \\
        Teegardens b & 0.03$_{0.01}^{0.10}$ & 54$_{18}^{204}$ \\
        Teegardens c & 0.40$_{0.16}^{1.13}$ & 1020$_{375}^{3207}$ \\
        TRAPPIST-1 d & 1.43$_{0.99}^{2.22}$ & 4240$_{2890}^{6775}$ \\
        TRAPPIST-1 e & 6.78$_{4.91}^{9.70}$ & 22263$_{16005}^{32437}$ \\
        TRAPPIST-1 f & 18.76$_{14.15}^{25.01}$ & 65213$_{48815}^{87970}$ \\
        \noalign{\smallskip}
        \hline
    \end{tabular}
    \begin{tablenotes}
        \item[] Median values for the volume-averaged shear modulus, $\mu$, and viscosity, $\eta$, are shown, along with the values of the first and third quartiles in the lower and upper indices, respectively.
    \end{tablenotes}
    \label{tab:app_rheo}
    \end{table}
    
\end{appendix}

\end{document}